\documentclass[reprint,pra,aps,nofootinbib]{revtex4-1}

\usepackage[utf8]{inputenc}
\usepackage{graphicx}
\usepackage{verbatim}
\usepackage{dsfont}
\usepackage{epstopdf}
\usepackage[caption=false]{subfig}
\usepackage{tikz}
\usetikzlibrary{calc}
\usepackage{relsize}
\usepackage[hidelinks]{hyperref}

\expandafter\let\csname equation*\endcsname\relax
\expandafter\let\csname endequation*\endcsname\relax
\usepackage{amsmath,amsfonts,amssymb,amstext}
\usepackage{cleveref}

\def\bra#1{\mathinner{\langle{#1}|}}
\def\ket#1{\mathinner{|{#1}\rangle}}

\begin{document}

\title[Complete classification of trapping coins for quantum walks on the 2D square lattice]{Complete classification of trapping coins for quantum walks on the 2D square lattice}

\author{B. Koll\'ar$^{1}$, A. Gily\'en$^{1,4,5}$, I. Tk\'{a}\v{c}ov\'{a}$^2$, T. Kiss$^1$, I. Jex$^3$ and M. \v{S}tefa\v{n}\'{a}k$^3$}
\affiliation{$^1$Wigner RCP, Konkoly-Thege M. u. 29-33, H-1121 Budapest, Hungary}
\affiliation{$^2$Department of Physics, V\v SB Technical University of Ostrava, 17. Listopadu 15, 70800 Ostrava-Poruba, Czech Republic}
\affiliation{$^3$Department of Physics, Faculty of Nuclear Sciences and Physical Engineering, Czech Technical University in Prague, B\v rehov\'a 7, 115 19 Praha 1 - Star\'e M\v{e}sto, Czech Republic}
%\affiliation{$^4$QuSoft, CWI, and University of Amsterdam, the Netherlands}
%\affiliation{$^5$Institute for Quantum Information and Matter, California Institute of Technology, USA}
\affiliation{$^4$QuSoft / CWI, Science Park 123, 1098 XG Amsterdam, the Netherlands}
\affiliation{$^5$Institute for Quantum Information and Matter, California Institute of Technology, 1200 E California Blvd, Pasadena, CA 91125, US}

\date{\today}

\begin{abstract}

One of the unique features of discrete-time quantum walks is called trapping, meaning the inability of the quantum walker to completely escape from its initial position, albeit the system is translationally invariant. The effect is dependent on the dimension and the explicit form of the local coin. A four state discrete-time quantum walk on a square lattice is defined by its unitary coin operator, acting on the four dimensional coin Hilbert space. The well known example of the Grover coin leads to a partial trapping, i.e., there exists some escaping initial state for which the probability of staying at the initial position vanishes. On the other hand, some other coins are known to exhibit strong trapping, where such escaping state does not exist. We present a systematic study of coins leading to trapping, explicitly construct all such coins for discrete-time quantum walks on the 2D square lattice, and classify them according to the structure of the operator and the manifestation of the trapping effect. We distinguish three types of trapping coins exhibiting distinct dynamical properties, as exemplified by the existence or non-existence of the escaping state and the area covered by the spreading wave-packet.
%We show that the area covered by the spreading walker can be well estimated by the overlap of two crossing ellipses. The parameters of the ellipses are defined by the properties of the coin.
\end{abstract}

\maketitle

%%%%%%%%%%%%%%%%%%%%%%%%%%%%%%%%%%%%%%
\section{Introduction}

Discrete time quantum walks \cite{Aharonov1993,Meyer1996} are non-trivial generalizations of discrete-time classical walks. These elementary constructs follow the rules of quantum mechanics and became versatile tools in various field of physics (for reviews see \cite{Kempe2003,Konno2008,reitzner2011quantum,Elias2012,Wang2013,portugal2013quantum}).
The motion of a single excitation in a solid state, the spreading of quantum information in a quantum network and even quantum computation can be modeled by quantum walks~\cite{Lovett2010}.

Recently, quantum walks have attracted  interest as simple quantum simulators, modeling the behavior of quantum particles under various conditions: the effect of decoherence \cite{Kendon2007,Annabestani2010}, electric fields \cite{Cedzich2013,Genske2013}, and percolation \cite{Romanelli2005,Kendon2010,Kollar2012,Kollar2014,Anishchenko2012,Darazs2013,Chandrashekar2014} were studied in detail. During the last decade, a number of state-of-the-art experiments \cite{OBrien2010,Zahringer2010,Schreiber2010,Schreiber2011, Schreiber2012,Kitagawa2012,Jeong2013,Wang2013,Genske2013,Jeong2013,Crespi2013,Robens2015,elster2015,xue2017,Nitsche2018,mataloni2019,xue2019} were performed validating the theoretical results and also benchmarking the achievable degree of quantum control and visibility.

Quantum walks serve as an elementary model for transport phenomena in physical systems. Spreading properties of quantum walks significantly differ from classical random walks. They can spread faster, thus speeding up random walk based search \cite{Childs2003,Ambainis2007,Szegedy2004,Ambainis2019}, leading to a number of possible applications in quantum information \cite{Magniez2006}. Nevertheless, there might be vertices which are almost never reached by the walker due to destructive interference, leading to infinite hitting times even for finite graphs \cite{krovi2006a,Krovi:2006}. However, for the initial vertex the expected return time is always finite for a finite graph, as follows from a general results for discrete-time unitary evolution  \cite{Grunbaum2013}. The expected return time to the exact initial state (state recurrence) is an integer \cite{Grunbaum2013}. This holds even for iterated open quantum evolution, provided it is described by a unital quantum channel \cite{Sinkovicz1}. The investigation was later extended to a broader class of iterated open quantum dynamics \cite{Sinkovicz2} and the result can be understood as a generalization of the Kac lemma \cite{Kac1947}. We note that in the case of subspace recurrence the expected return time is a rational number \cite{gruenbaum2014}.

Quantum walks are known for their typical ballistic spreading \cite{Ambainis2001}. However, for a quantum walk on a two-dimensional lattice there exist some coins which lead to limited spreading for some initial states. In particular, for a Grover coin one can observe a probability peak situated at the origin of the walk, discovered by Mackay, Bartlett, Stephenson and Sanders \cite{Mackay2002}. We will refer to this property as trapping. Let us note that the term localization is sometimes used for the same effect in the literature, however localization \cite{Keating2007} is often used in a different context, e.g., in Anderson localization -- a phenomenon arising from spatial randomness \cite{Torma2002,Schreiber2011,Crespi2013,rakovszky2015localization,Edge2015,Joye2010,Joye2012,Ahlbrecht2011,Ahlbrecht2012}, exponential localization of topologically protected states \cite{Kitagawa2012,Kitagawa2010,Asboth2012}, or oscillatory localization~\cite{ambainis2016OscillatoryLoc}. The effect of trapping by a Grover coin for discrete-time quantum walks on 2D integer lattice was rigorously proven by Inui, Konishi and Konno \cite{Inui2004}. Implications of trapping for stationary measures of quantum walks were discussed in \cite{machida2015,komatsu2017}. We note that trapping is not limited to the square lattice, but can be found in any $d$-dimensional lattice.  For quantum walks on a line non-trivial trapping coins need to have at least three-dimensions~\cite{Inui2005}.
Trapping coins of dimensions greater than 3 were also identified \cite{inui2005b} and further studied in \cite{miyazaki2007,Bezdekova2015}. Several extensions of the three-state Grover coin featuring trapping were introduced \cite{stefanak2012} and investigated in detail \cite{falkner2014,stefanak2014,machida2015,Ko2016}. A full classification of three-dimensional coins leading to trapping for a quantum walk on a line was provided in \cite{Stefanak2013}. Likewise, trapping effect on a 2D integer lattice is not limited to the Grover coin: a family of coins with this property was introduced by Watabe, Kobayashi, Katori and Konno \cite{Watabe2008}. A systematic search for coins exhibiting trapping revealed that an even stronger type of trapping exist: it is possible that all initially localized states remain trapped  \cite{Kollar2015}. Although \cite{Kollar2015} presented a multiple-parameter class of coins exhibiting one or the other type of trapping, a complete classification was lacking.

In this paper we construct and classify all trapping coin operators for a discrete-time quantum walk on a 2D integer lattice, based on the observation that the localized eigenstates of the walk have a finite support -- in fact involving only four lattice sites. We classify trapping coins according to the possible dynamical behavior of the walk, with respect to a walker starting from a single vertex. For the first class of coins there always exists a trapped component, while the spreading part of the wave function is approximately present in an area characterized by a cross-section of two distinct ellipses. The form of the ellipses can be determined from the parameters of the coin. For the second class of coins there exists a unique escaping initial state which does not remain trapped. The characteristic spreading pattern is also formed by a cross-section of two ellipses, however, in this case the two ellipses may coincide. For the last type of trapping coins the escaping states form a two-dimensional subspace and the walk dynamics is essentially one-dimensional.

The paper is organized as follows. In section \ref{Sec:II} we define our model and introduce the effect of trapping. Section \ref{Sec:Construction} focuses on the action of the evolution operator on the stationary state. We derive two mutually exclusive conditions, one of which the trapping coin has to fulfill. The investigation of these two cases is the subject of section \ref{Sec:MainRes}, where we derive the explicit form of the trapping coin operators. 
The properties of the coin classes are investigated in Section~\ref{sec:5}. We focus on the existence and uniqueness of the escaping state and the area covered by the spreading part of the walk.
We summarize our results in Section~\ref{sec:6}. Finally, in Appendix \ref{App:support} we prove that the localized states can be decomposed into eigenstates supported on two by two regions of the lattice.

%%%%%%%%%%%%%%%%%%%
\section{Model}
\label{Sec:II}

Let us consider a four-state discrete-time quantum walk on a two-dimensional square lattice. The Hilbert space of the walk can be decomposed as
	\begin{equation}
	\mathcal{H} = \mathcal{H}_P \otimes \mathcal{H}_C\,,
	\end{equation}
where $\mathcal{H}_P$ is the position space spanned by the orthonormal set $\{| x,y \rangle\}$ with $x,y \in \mathbb{Z}$ indexing the positions on the lattice. The coin space $\mathcal{H}_C$ is spanned by the orthonormal basis defining possible movements of the particle to the left $ | L \rangle,$ down $ | D \rangle,$ up $ | U \rangle$ and right $ | R \rangle.$  
A single step of the time evolution is generated by the unitary operator
	\begin{equation}
 	\hat{U} = \hat{S} \cdot \left( \hat I_P \otimes \hat{C} \right)\,.
	\end{equation}
Here $\hat{S}$ is the shift operation responsible for the conditional displacement which is defined by its action on the basis states
	\begin{align*}
 	\hat{S} | x,y\rangle | L \rangle & =   | x-1,y\rangle| L \rangle \,,  \quad
 	\hat{S} | x,y\rangle| D \rangle = | x,y-1\rangle| D \rangle \,, \\
	\hat{S} | x,y\rangle| U \rangle & =  | x,y+1\rangle| U \rangle \,, \quad
	\hat{S} | x,y\rangle| R \rangle =  | x+1,y\rangle| R \rangle\,.
	\end{align*}
$\hat I_P$ is the identity on the position space. Finally, $\hat{C}$ is the unitary coin operator acting only on the coin space $\mathcal{H}_C$ and mixing the coin states in the following way
 	\begin{equation}
 	\label{Eq:CoinAct} 
 	\hat{C}|j\rangle  = \sum_{i,j} C_{ij}|i\rangle
    , \quad i,j\in \{L, D, U, R\}.
 	\end{equation}
The matrix representation $C_{ij}$ of the operator $\hat{C}$ in the standard basis $|L\rangle, |D\rangle, |U\rangle, |R\rangle$ will be referred to as the coin $C$. We emphasize that throughout the paper we use the indices  $L, D, U, R$ for rows and columns of the coin $C.$ For instance, matrix element $C_{RU}$ corresponds to $C_{43}.$ 

We consider initial states residing on a single vertex, that we identify with the origin of the lattice without loss of generality. We still have the freedom to choose the initial coin state  $|\psi_C\rangle\in{\mathcal H}_C$, i.e. the complete form of the starting state of the walk is given by  
$$
|\psi(0)\rangle = |0,0\rangle|\psi_C\rangle .
$$
The discrete-time evolution of the walk is given by repeating the evolution operator on the initial state
	\begin{equation*}
	|\psi(t)\rangle = \hat{U}^t |\psi(0)\rangle. 
	\end{equation*}
The state of the walk after $t$ steps can be decomposed into the standard basis according to
$$
|\psi(t)\rangle = \sum_{x,y} \sum_{i} \psi_i (x,y,t) |x,y\rangle |i\rangle 
$$
where $\psi_i(x,y,t)$, $i\in\{L,D,U,R\}$ are the amplitudes of the particle at position $(x,y)$ with coin state $i$. The probability distribution on the square grid is given by
\begin{align*}
P(x,y,t) & =    |\psi_L(x,y,t)|^2 + |\psi_D(x,y,t)|^2 + \\
 &\phantom{=}  +|\psi_U(x,y,t)|^2 + |\psi_R(x,y,t)|^2.
\end{align*}

Now we turn to the trapping of quantum walks on a square grid, which is the central topic of our paper. We say that a quantum walk operator is trapping if there is an initial coin state $\ket{\psi_C}$ such that the long-time average probability of finding the walker at the initial position is non-vanishing, i.e., 
\begin{equation}
\lim_{T\to \infty} \frac{1}{T}\sum_{t=1}^{T} P(0,0,t)\geq p>0. 
%\frac{1}{T}\sum_{t=1}^{T}\left\lVert\bra{0,0}\otimes \bra{i} \hat{U}^t \ket{0,0}\otimes\ket{\phi} \right\rVert^2 \geq p>0.
\end{equation}
It was observed that under cyclic boundary conditions, trapping requires a highly degenerate spectrum, featuring flat bands~\cite{Inui2004}. This result was later extended, showing that for a quantum walk on an infinite lattice a coin operator can be trapping (if and) only if the evolution operator $\hat U$ has an infinitely degenerate eigenvalue~\cite{Tate2017}. 

In the following we construct trapping coins based on the properties of eigenstates corresponding to the degenerate eigenvalue.
Since the global phase is irrelevant, we assume without loss of generality that $1$ is a degenerate eigenvalue, so we will work with solutions~of
	\begin{align}\nonumber\\[-6.5mm]
	\label{Eq:UxyCharEq}
	\hat{U} |\psi_{st}\rangle = \hat{S}(\hat I_P \otimes \hat{C}) |\psi_{st}\rangle = |\psi_{st}\rangle. 
	\end{align}
In Appendix \ref{App:support} we prove that the corresponding eigenstates can be chosen such that they have support of size (at most) $2\times 2$ on the lattice. Then a stationary eigenstate occupying vertices $(x,y)$, $(x,y+1)$, $(x+1,y)$ and $(x+1,y+1)$ can be written in the form
\begin{align}\label{Eq:StStatexy}
\nonumber	|\psi_{st}^{(x,y)}\rangle = & \,\, |x,y\rangle|\xi^{(0,0)}\rangle + |x,y+1\rangle|\xi^{(0,1)}\rangle + \\
 & + |x+1,y\rangle|\xi^{(1,0)}\rangle + |x+1,y+1\rangle|\xi^{(1,1)}\rangle. 
\end{align}	
Here $|\xi^{(i,j)}\rangle$, $(i,j\in \{0,1\})$ denote the local coin states which are in general given by
\begin{equation}
\label{Eq:XiCoef}
|\xi^{(i,j)}\rangle =\xi_L^{(i,j)}|L\rangle + \xi_D^{(i,j)}|D\rangle + \xi_U^{(i,j)}|U\rangle + \xi_R^{(i,j)}|R\rangle.
\end{equation}
Due to the translational invariance of the considered walk the local coin states $|\xi^{(i,j)}\rangle$ are independent of $(x,y)$. Hence, the stationary states $|\psi_{st}^{(x,y)}\rangle$ have the same form for all $(x,y)$, only their support on the lattice is different. Therefore, it is sufficient to consider only one of the stationary states e.g.~$|\psi_{st}^{(0,0)}\rangle$.
(We remark that due to chiral symmetry, every eigenstate $\sum_{x,y}\ket{x,y}\ket{\xi^{(x,y)}}$ has a chiral counterpart $\sum_{x,y}(-1)^{x+y}\ket{x,y}\ket{\xi^{(x,y)}}$ and the corresponding eigenvalues differ by a factor of $(-1)$~\cite{Stefanak2010}.)

In the following section we study the structure of the stationary state $|\psi_{st}^{(0,0)}\rangle$ based on \cref{Eq:UxyCharEq} in order to later find all trapping coins of the four-state discrete-time quantum walks on the two-dimensional lattice.

%%%%%%%%%%%%%%%%%%%%%%%%%%%%%
\section{Restrictions on the amplitudes of trapped eigenstates}
\label{Sec:Construction}

Our first task in this section is to determine the possible values of the 16 coefficients $\xi_i^{(m,n)}$ in equation (\ref{Eq:StStatexy}).
It turns out that some of these coefficients have to be zero. Let us examine the action of the inverse shift $\hat S^{-1}$ on the stationary state $|\psi_{st}^{(0,0)}\rangle$. From equation (\ref{Eq:UxyCharEq}) we have
	\begin{equation}
	\label{Eq:SInvEqualC}
	(\hat I_P \otimes \hat{C}) |\psi_{st}^{(0,0)}\rangle = \hat{S}^{-1}|\psi_{st}^{(0,0)}\rangle.
	\end{equation}
The left-hand side of this equation changes the coin states without touching their positions. On the other hand, the right hand side changes only the positions. This equality cannot hold if $\hat S^{-1}$ steps out of the given $2\times 2$ region. This eliminates half of the coefficients defining the local coin states (\ref{Eq:XiCoef}) of the general stationary state (\ref{Eq:StStatexy}). For notational convenience we will denote the remaining, potentially non-zero coefficients, by $a=\xi^{(0,0)}_L,b=\xi^{(0,0)}_D, c=\xi^{(0,1)}_L, d=\xi^{(0,1)}_U, e=\xi^{(1,0)}_D, f=\xi^{(1,0)}_R, g=\xi^{(1,1)}_U, h=\xi^{(1,1)}_R$, i.e. the local coin states have the form
\begin{align}\nonumber\\[-9mm]
\nonumber |\xi^{(0,0)}\rangle & =  a|L\rangle + b|D\rangle, \\
\nonumber |\xi^{(0,1)}\rangle & =  c|L\rangle + d|U\rangle, \\
\nonumber |\xi^{(1,0)}\rangle & =  e|D\rangle + f|R\rangle, \\
 |\xi^{(1,1)}\rangle & =  g|U\rangle + h|R\rangle. \label{local:coin:states}
\end{align}
As illustrated by Figure~\ref{fig:eigen}, in order to fulfill \cref{Eq:SInvEqualC} the coin operator $\hat{C}$ has to act on the local coin states $|\xi^{(m,n)}\rangle$ in the following way:
\begin{align}
\nonumber \hat{C} |\xi^{(0,0)}\rangle & = d |U\rangle + f |R\rangle,\\ 
\nonumber \hat{C} |\xi^{(0,1)}\rangle & = b |D\rangle + h |R\rangle, \\
\nonumber \hat{C} |\xi^{(1,0)}\rangle & = a |L\rangle + g |U	\rangle, \\ 
\hat{C} |\xi^{(1,1)}\rangle & = c |L\rangle + e |D\rangle.  \label{Eq:Caction}
\end{align}	
The relations (\ref{Eq:Caction}) can be written in a matrix form as 
	\begin{align}
	\label{eq:Cimplicit}
	C\cdot
	\underset{A}
	{\underbrace{\left(\begin{array}{cccc}
				a & c & 0 & 0\\
				b & 0 & e & 0\\
				0 & d & 0 & g\\
				0 & 0 & f & h
			\end{array}\right)}}
	=
	\underset{B}
	{\underbrace{\left(\begin{array}{cccc}
				0 & 0 & a & c\\
				0 & b & 0 & e\\
				d & 0 & g & 0\\
				f & h & 0 & 0\\
			\end{array}\right)}},
	\end{align}
where $C$ is the specific coin matrix we are searching for and the individual columns of the matrices $A, B$ represent the vectors on the left hand side and the right hand side in (\ref{Eq:Caction}). The 16 individual equations can be considered as detailed balance conditions between the amplitudes of the stationary state. Moreover, the matrix $C$ has to be unitary, i.e. $C^\dagger C=I$. This leads us to the following relation for the matrices $A$ and $B$
$$
A^\dagger A - B^\dagger B = A^\dagger C^\dagger C A-B^\dagger B = B^\dagger B-B^\dagger B = 0,
$$
which can be written in a matrix form as
\begin{widetext}
\begin{align} \nonumber\\[-10mm]\label{eq:unitarity}
	\left(
	\begin{array}{cccc}
		|a|^2+|b|^2-|d|^2-|f|^2 & c a^*-h f^* & e b^*-g d^* & 0 \\
		a c^*-f h^* & |c|^2+|d|^2-|b|^2-|h|^2 & 0 & g d^*-e b^* \\
		b e^*-d g^* & 0 & |e|^2+|f|^2-|a|^2-|g|^2 & h f^*-c a^* \\
		0 & d g^*-b e^* & f h^*-a c^* & |g|^2+|h|^2-|c|^2-|e|^2 \\
	\end{array}
	\right)	= 0.\\[-8mm]\nonumber
\end{align}
\newpage
\begin{figure}[t]
	\centering
	\hspace{12pt}
	\includegraphics[width=0.35\textwidth]{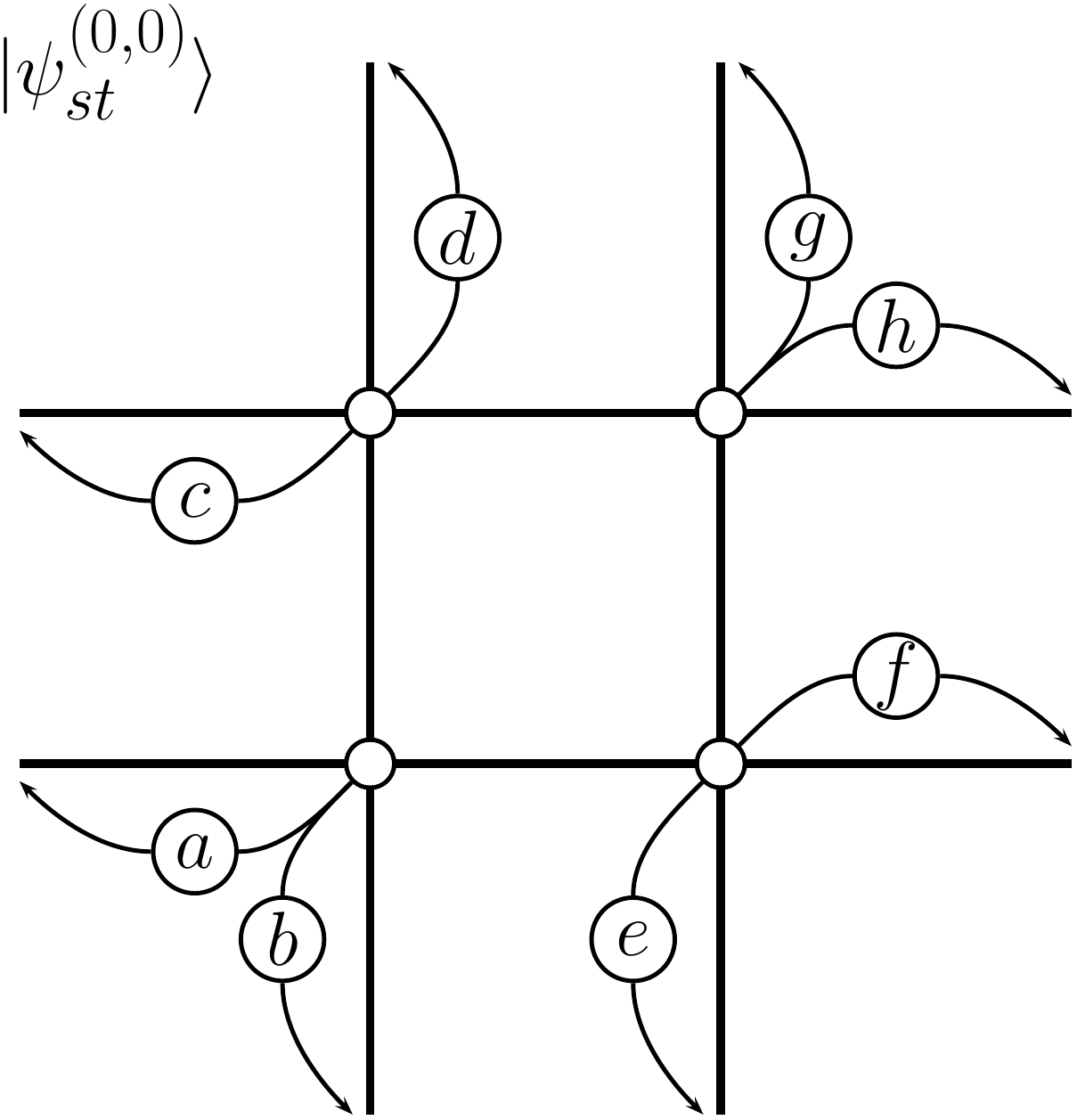}\hfill
	\includegraphics[width=0.35\textwidth]{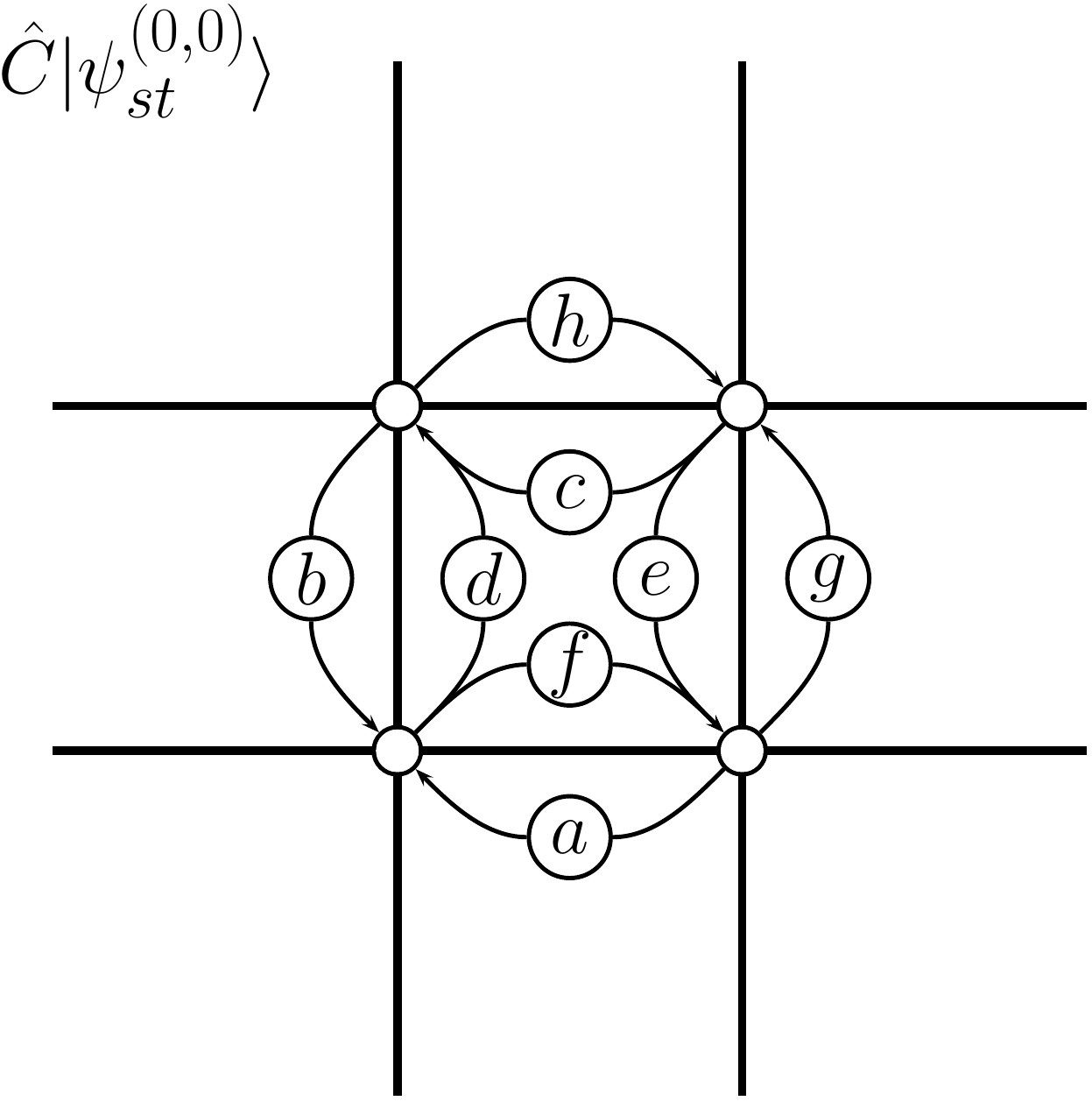}
	\hspace{12pt}
	\caption{On the left we display a schematic representation of the stationary state $\ket{\psi^{(0,0)}_{st}}$. The letters in circles denote the amplitudes of the respective local coin states (\ref{local:coin:states}). On the right we show the stationary state after the application of the coin operator $\hat C$, which acts on the local coin states according to (\ref{Eq:Caction}). The step operator $\hat S$ propagates the amplitudes in the direction of the arrow, thus returning the state $\hat C\ket{\psi^{(0,0)}_{st}}$ back to $\ket{\psi^{(0,0)}_{st}}$. } 
	\label{fig:eigen}
\end{figure}
\noindent After removing redundant equations from \eqref{eq:unitarity} one can see that $A^\dagger A-B^\dagger B = 0$ is equivalent to the following set of equations:
\begin{align}
	|a|^2 + |b|^2 =\, & |d|^2 + |f|^2 \label{eq:C1}\\
	|g|^2 + |h|^2 =\, & |c|^2 + |e|^2 \label{eq:C4}\\
	|c|^2 + |d|^2 =\, & |b|^2 + |h|^2 \label{eq:C2}\\ 
	%|e|^2 + |f|^2 =\, & |a|^2 + |g|^2 \label{eq:C3}  
	a c^* =\, & f h^*     \label{eq:acconj}\\
	b e^* =\, & d g^*\!.  \label{eq:beconj}
\end{align}

Equation \eqref{eq:acconj} implies $|ac| = |fh|$ and \eqref{eq:beconj} implies $|be| =|dg|$, which tells:
\begin{align}
	|c|^2 + |d|^2 =& |b|^2 + |h|^2 &
	|c|^2 + |d|^2 =& |b|^2 + |h|^2 \nonumber\\
	\Downarrow & \cdot|a|^2 &
	\Downarrow & \cdot|g|^2 \nonumber\\
	|a|^2|c|^2 + |a|^2|d|^2 =& |a|^2|b|^2 + |a|^2|h|^2 &
	|g|^2|c|^2 + |g|^2|d|^2 =& |g|^2|b|^2 + |g|^2|h|^2 \nonumber\\
	\Updownarrow & (|ac| = |fh|) &
	\Updownarrow & (|be| = |dg|) \nonumber\\
	|f|^2|h|^2 + |a|^2|d|^2 =& |a|^2|b|^2 + |a|^2|h|^2 &
	|g|^2|c|^2 + |e|^2|b|^2 =& |g|^2|b|^2 + |g|^2|h|^2 \nonumber\\
	\Updownarrow & &
	\Updownarrow & \nonumber\\
	|h|^2\left(|f|^2-|a|^2\right) =& |a|^2\left(|b|^2-|d|^2\right) &
	|g|^2\left(|c|^2-|h|^2\right) =& |b|^2\left(|g|^2-|e|^2\right) \nonumber\\
	\Updownarrow & (\text{by } \eqref{eq:C1}) &
	\Updownarrow & (\text{by } \eqref{eq:C4}) \nonumber\\
	\left(|h|^2-|a|^2\right)\left(|f|^2-|a|^2\right)=&0 &
	\left(|g|^2-|b|^2\right)\left(|g|^2-|e|^2\right)=&0 \nonumber\\
	\Updownarrow &\, (\text{by } \eqref{eq:C1}-\eqref{eq:C2}) &
	\Updownarrow &\, (\text{by } \eqref{eq:C1}-\eqref{eq:C2}) \nonumber\\
	 \left\{
	\begin{array}{lrl}
		\textbf{I)}& \text{ either } & |a|=|h| \text{ and } |c|=|f| \\
		\hline
		\textbf{II)} & \text{or}   & |a|=|f| \text{ and } |b|=|d| \\
		     &\text{and}    & |c|=|h| \text{ and } |g|=|e| \\
	\end{array}
	\right. \kern-25mm& &
	\phantom{--}  \left\{
	\begin{array}{lrl}
		\textbf{I)} & \text{either} & |g|=|b| \text{ and } |d|=|e| \\
		\hline
		\textbf{II)} & \text{or}     & |g|=|e| \text{ and } |c|=|h| \\
		            &\text{and}    & |b|=|d| \text{ and } |a|=|f| \\
	\end{array}
	\right. \kern-25mm& \nonumber
\end{align}

Since case {\bf II)} on the left- and right-hand sides coincide, we are left with two possibilities:
\begin{align}
    \text{either \bf{II}) } |a|=|f| \text{ and } |c|=|h| 
	&\text{ and } |b|=|d| \text{ and } |g|=|e|, \label{eq:new} \\
	\text{or \bf{I}) } |a|=|h| \text{ and } |c|=|f| 
	&\text{ and } |g|=|b| \text{ and } |d|=|e| \label{eq:balint} \text{ (while {\bf II} does not hold).}    
\end{align}
\newpage
\end{widetext}

Now we can proceed using case separation based on which one of the two sets of equations, \eqref{eq:new} or \eqref{eq:balint}, holds. 
As we will show, these two cases correspond to whether $\det A=b c f g - a d e h$ is zero or not.

%%%%%%%%%%%%%%%%%%%%%%%%%%%%%
\section{Classification of trapping coins}
\label{Sec:MainRes}

At the end of the previous section we have derived two mutually exclusive conditions \eqref{eq:new}-\eqref{eq:balint}, which have to be fulfilled for trapping coins. Based on these conditions we can construct all different types of trapping coins determined by the rank of the matrix $A$. In order to provide a full classification, we briefly study degenerate cases as well, even if they lead to trivial dynamics.

\subsection{Case I: \texorpdfstring{$\det A$}{det A} is non-zero}
\label{Subsec:STtrapp}

This part is devoted to the description of trapping coins corresponding to eigenstates satisfying \cref{eq:balint}. We refer to this class of coins as Type ${\bf I}$. As we will see, in this case $\det A\neq 0$ so the matrix $A$ has full rank. Hence, there exists an inverse matrix $A^{-1}$ and thus $C$ is uniquely determined by the amplitudes of $\ket{\psi_{st}^{(0,0)}}$ via \cref{eq:Cimplicit} as follows:
\begin{equation}
C = B A^{-1}.
\end{equation}

Let us now turn to a particular parametrization of the amplitudes. We assume, without loss of generality, that the norm of the stationary state $\ket{\psi_{st}^{(0,0)}}$ is $2$. Together with conditions \eqref{eq:C1}-\eqref{eq:C4} and \eqref{eq:balint} this implies 
$$
|a|^2+|b|^2=|d|^2+|f|^2 = 1 . 
$$
Therefore, we can write the magnitudes $|a|,|b|,|d|,|f|$ as sine and cosine functions 
\begin{align}
\nonumber |a| & = \sin\delta_1 = |h| , \quad |b| = \cos\delta_1 = |g|, \\
	|c| & = \sin\delta_2 = |f| , \quad |d| = \cos\delta_2 = |e|, \label{eq:stMagnitues}
\end{align}
where $\delta_1,\delta_2\in [ 0,\pi/2 ]$. Note that $\delta_1\neq\delta_2$, since we assume that \eqref{eq:new} does not hold. It is easy to see that this parametrization implies $|b c f g|\neq |a d e h|$ and therefore $\det A$ is indeed non-zero.

Now we also parameterize the phases of the amplitudes $a, \dots , h$. For an $x\in \mathbb{C}$ let $\phi_x\in\mathbb{R}$ denote its phase, such that $x=|x|e^{i\phi_x}$. Equation \eqref{eq:acconj} shows that we can assume, without loss of generality, that $\phi_a-\phi_c=\phi_f-\phi_h.$ Similarly by \eqref{eq:beconj} we have that $\phi_b-\phi_e=\phi_d-\phi_g$. Since we can arbitrarily choose the global phase of the stationary state, we also assume $\phi_a=0$. If some of the parameters are $0$ then some phases become irrelevant, nevertheless the above assumptions do not break generality. Thus the amplitudes can be parameterized as follows:
\begin{align}
a =& s_1, & b =& c_1 e^{i\left(\phi_d+\phi_e-\phi_g\right)} , \nonumber \\
c =& s_2 e^{i\left(\phi_h-\phi_f\right)} , & d =& c_2 e^{i\phi_d} , \nonumber \\
e =& c_2 e^{i\phi_e} , & f =& s_2 e^{i\phi_f} , \nonumber \\
g =& c_1 e^{i\phi_g} , & h =& s_1 e^{i\phi_h} ,
\label{Eq:paramFinalI}
\end{align}
where $s_i = \sin\delta_i,\, c_i = \cos\delta_i$ for $i\in\{1,2\}$ and $\delta_1\neq\delta_2$.
Then the explicit form of Type ${\bf I}$ coins is given by
\begin{widetext}
\begin{equation}
	\label{Cst}
C_{\bf I} = B A^{-1} =  \left( \begin{array}{rrrr}
		-e^{i (\phi_d-\phi_g)} c_1 c_2 &
		e^{-i \phi_e} s_1 c_2 &
		\,\,\,e^{i (\phi_h-\phi_f-\phi_g)} c_1 s_2 &
		e^{-i \phi_f}  s_1 s_2 \\
		e^{i (\phi_d+\phi_e+\phi_f-\phi_g-\phi_h)} c_1 s_2 &
		-e^{i (\phi_f-\phi_h)} s_1 s_2 &
		e^{i (\phi_e-\phi_g)} c_1 c_2 &
		e^{i (\phi_e-\phi_h)} s_1 c_2 \\
		e^{i \phi_d} s_1 c_2 &
		e^{i (\phi_g-\phi_e)} c_1 c_2 &
		- e^{i (\phi_h-\phi_f)} s_1 s_2 &
		e^{i(\phi_g - \phi_f)} c_1 s_2 \\
		e^{i \phi_f} s_1 s_2 &
		\,\,\,e^{i (\phi_f+\phi_g-\phi_d-\phi_e)} c_1 s_2 &
		e^{i (\phi_h-\phi_d)} s_1 c_2 &
		\,\,-e^{i (\phi_g-\phi_d)} c_1 c_2 
	\end{array} \right).
\end{equation}
%%%%%%%%%%%%%%%%%%%%%%%%%%%%%%%%%%%%%%%%%%%%%%
%%%%%%%%%% Checked by Mathematica: %%%%%%%%%%%
\begin{comment}
$Assumptions = Element[d1 | d2 | fd | fe | ff | fg | fh, Reals];
s1 = Sin[d1];
c1 = Cos[d1];
s2 = Sin[d2];
c2 = Cos[d2];
(CI = {{-Exp[I (fd - fg)] c1 c2, Exp[-I fe] s1 c2, Exp[I (fh - ff - fg)] c1 s2, 
Exp[-I ff] s1 s2}, {Exp[I (fd + fe + ff - fg - fh)] c1 s2, -Exp[I (ff - fh)] s1 s2, Exp[I (fe - fg)] c1 c2, Exp[I (fe - fh)] s1 c2}, {Exp[I fd] s1 c2, Exp[I (fg - fe)] c1 c2, -Exp[I (fh - ff)] s1 s2, Exp[I (fg - ff)] c1 s2}, {Exp[I ff] s1 s2, Exp[I (ff + fg - fd - fe)] c1 s2, Exp[I (fh - fd)] s1 c2, -Exp[I (fg - fd)] c1 c2}}) // MatrixForm
Det[CI] // Simplify
(DI = CI - DiagonalMatrix[{x, y, 1/y, 1/x}]) // MatrixForm;
Det[DI] // Simplify
Det[DI[[1 ;; 3, 1 ;; 3]]]/x - Det[DI[[2 ;; 4, 2 ;; 4]]]*x // Simplify
Det[DI[[{1, 2, 4}, {1, 2, 4}]]]/y - Det[DI[[{1, 3, 4}, {1, 3, 4}]]]*y // Simplify
\end{comment}
%%%%%%%%%%%%%%%%%%%%%%%%%%%%%%%%%%%%%%%%%%%%%%
\end{widetext}
\noindent The corresponding stationary states come in chirally symmetric pairs that are proportional to
\begin{align}
\phantom{+} & \ket{0,0}\big(s_1 \ket{L} + c_1 e^{i\left(\phi_d+\phi_e-\phi_g\right)}\ket{D}\big) + \nonumber\\
\pm & \ket{0,1}\big(s_2 e^{i\left(\phi_h-\phi_f\right)}    \ket{L} + c_2 e^{i\phi_d}\ket{U}\big) +\nonumber\\
\pm & \ket{1,0}\left(c_2 e^{i\phi_e} \ket{D} + s_2 e^{i\phi_f}\ket{R} \right) + \nonumber\\
+ & \ket{1,1}\left(c_1 e^{i\phi_g} \ket{U} + s_1 e^{i\phi_h}\ket{R}\right), \nonumber
\end{align}
so the probability distribution of the stationary states is uniform across the $2\times 2$ unit cell:
\begin{align*}
P(0,0) = P(0,1) = P(1,0) = P(1,1) = \frac{1}{4}.
\end{align*}
In Section~\ref{sec:5} we identify the degenerate cases $\delta_1,\delta_2\in\{0,\frac\pi2\}$; the degeneracy leads to two additional stationary states, but they only differ by some complex phases.
As an example, consider the case $\delta_1 = \frac{\pi}{2}, \delta_2 = 0$. The coin $C_{\bf I}$ then becomes
\begin{equation*}
\left( \begin{array}{cccc}
0 &	e^{-i \phi_e} &	0 &	0 \\
0 &	0 & 0 &	e^{i (\phi_e-\phi_h)} \\
e^{i \phi_d} & 0 & 0 & 0\\
0 & 0 &	e^{i (\phi_h-\phi_d)} & 0 
\end{array} \right),
\end{equation*}
and the stationary states become 
\begin{align*}
\ket{0,0}\!\ket{L} 
+ \frac{e^{i\phi_d}}{\lambda}\ket{0,1}\!\ket{U} 
+ \frac{e^{i\phi_h}}{\lambda^2}\ket{1,1}\!\ket{R}
+ \frac{e^{i\phi_e}}{\lambda^3}\ket{1,0}\!\ket{D},
\end{align*}	
corresponding to eigenvalues $\lambda\in\{1,i,-1,-i\}$.
The other case $\delta_1=0, \delta_2 = \frac{\pi}{2}$ is analogous.

%%%%%%%%%%%%%%%%%%%%%%%%%%%%%%%%%%%%%%%%%%%%%%
%%%%%%%%%% Checked by Mathematica: %%%%%%%%%%%
\begin{comment}
(mat = {{0, Exp[-I fe], 0, 0}, {0, 0, 0, Exp[I (fe - fh)]}, {Exp[I fd], 0, 0, 0}, {0, 0, Exp[I (fh - fd)], 0}}) // MatrixForm
vec = {1, Exp[I fe]/l^3, Exp[I fd]/l, Exp[I fh]/l^2}
Table[(mat.vec /. {l -> I^k} // Simplify)/vec /. {l -> I^k}, {k, 0, 3}]
\end{comment}
%%%%%%%%%%%%%%%%%%%%%%%%%%%%%%%%%%%%%%%%%%%%%%

\subsection{Case {\bf II}: \texorpdfstring{$\det A$}{det A} is zero}
\label{Subsec:detAzero}

The remaining cases correspond to \cref{eq:new} which describes the situation when $\det A = 0$.  %, i.e. $A$ is not full rank. 
To see this, one can multiply the two equations \eqref{eq:acconj} and \eqref{eq:beconj} to obtain
$$adc^*g^*=bfe^*h^*.$$ 
Multiplying both sides with $cgeh$ we get the equality 
$$
adeh|cg|^2=bcfg|eh|^2 .
$$
Due to \eqref{eq:new} this is further equivalent to 
$$
0=|eh|^2\underbrace{(bcfg-adeh)}_{\det A},
$$ 
which implies that $e$ or $h$ or $\det A=0.$ If one of the parameters $e$ or $h$ is equal to zero, due to equation (\ref{eq:new}) also $g$ or $c$ equals zero which results in $\det A =0$ as well.

This case can be further divided to 2 sub-cases depending on the rank of the matrix $A$ which can be three or (at most) two. 

Before separating these cases we introduce a convenient parametrization of the amplitudes that naturally fits case {\bf II)}. Similarly to case {\bf I)}, we assume without loss of generality that the norm of $\ket{\psi_{st}^{(0,0)}}$ is $\sqrt{2}$, which together with \cref{eq:new} enforces
$$
|a|^2+|b|^2+|c|^2+|e|^2 = 1.
$$
We can also assume without loss of generality that the phase of the parameter $a$ is zero, thus $a\in\mathbb{R}$. All such amplitudes $a, \dots, h$ satisfying \cref{eq:acconj,eq:beconj,eq:new} can be parameterized (choosing $\sin^2(\delta_1):=|a|^2+|c|^2$) as 
\begin{align}
a =& s_1 s_3 & 
b =& c_1 s_2 e^{i\left(\phi_d+\phi_e-\phi_g\right)} \nonumber \\
c =& s_1 c_3 e^{i\left(\phi_h-\phi_f\right)} & 
d =& c_1 s_2 e^{i\phi_d} \nonumber \\
e =& c_1 c_2 e^{i\phi_e} & 
f =& s_1 s_3 e^{i\phi_f} \nonumber \\
g =& c_1 c_2 e^{i\phi_g} & 
h =& s_1 c_3 e^{i\phi_h},
\label{Eq:paramNew}
\end{align}
where $\delta_k\in[0,\pi/2]$, $s_k=\sin \delta_k$ and $c_k=\cos \delta_k$ for $k=1,2,3$. 

\subsubsection{Case IIa: matrix \texorpdfstring{$A$}{A} has rank 3}

Let us now consider the situation when $A$ has rank~3. We refer to the corresponding class of coins as Type {\bf IIa}. In this case $A$ is not invertible and the amplitudes only determine the coin up to a single ``phase'' parameter, in contrast to {\bf I)}. As before, our starting point is \cref{eq:Cimplicit}, which implies that $B$ must also have rank 3. So we can find vectors $v_A$ and $v_B$ such that $\lVert v_A\rVert=\lVert v_B\rVert>0$ and
$v_A A=0$, $v_B B=0$.
%\begin{equation}%\label{va:vb}
%v_A A=0,\quad v_B B=0.
%\end{equation}
Since $C$ maps the orthogonal complement of the column space of $A$ to the orthogonal complement of the column space of $B$, we must have that for some $\eta\in(-\pi,\pi]$
\begin{align}\nonumber \\[-8mm]\label{CIIa:va:vb}
C v_A^\dagger = e^{i \eta}v_B^\dagger.  \\[-3mm]\nonumber
\end{align}
Since $A$ has rank 3, at least one of its columns must be linearly dependent from the other columns. This column is therefore redundant in \cref{eq:Cimplicit}, and it can be removed. Instead of removing this column we replace it with equation \cref{CIIa:va:vb}, resulting in the new equation
\begin{equation}	\label{CIIa:1}
	C\tilde{A} =\tilde{B},
\end{equation} 
where $\tilde{A}$ is the full-rank matrix obtained from $A$ by replacing one redundant column by $v_A^\dagger$, and similarly $\tilde{B}$ is obtained by replacing the corresponding column in $B$ with $e^{i \eta}v_B^\dagger$. Therefore, we can describe the Type {\bf IIa} solutions in the form
\begin{equation}\label{CIIa:2}
C_{\bf IIa}=\tilde{B}\tilde{A}^{-1}.
\end{equation}

Now we explicitly construct $C_{\bf IIa}$, first assuming $\delta_1\in(0,\pi/2)$ and $\delta_2,\delta_3\in[0,\pi/2)$. In this case the last three columns of $A$ are linearly independent, and we can chose 
\begin{align}
v_A & := (d e h, -c f g, -c e h, c e g), \label{eq:presc}\\
v_B & := (e g h, -c g h, -a e h, b c g). \nonumber 
\end{align}
\vskip0.3mm

We can then replace the first columns, yielding
\begin{equation}
	\label{Eq:tildeAB}
		\tilde{A} = 
			\left(
			\begin{array}{rrrr}
			    (deh)^{*} & c & 0 & 0\\
	\kern-2.3mm-(cfg)^{*}& 0 & e & 0\\
	\kern-2.3mm-(ceh)^{*} & d & 0 & g\\
			    (ceg)^{*} & 0 & f & h
			\end{array}
			\right)\!\!, \,\,\,
		\tilde{B} = 			\left(
			\begin{array}{rrrr}
			    e^{i\eta}(egh)^{*} & 0 & a & c\\
	\kern-2.3mm-e^{i\eta}(cgh)^{*} & b & 0 & e\\
	\kern-2.3mm-e^{i\eta}(aeh)^{*} & 0 & g & 0\\
			    e^{i\eta}(bcg)^{*} & h & 0 & 0
			\end{array}
			\right)\!\!.
\end{equation}
\begin{widetext}
Using the parametrization of \cref{Eq:paramNew} and setting $\Xi := \left(e^{i \eta}-1\right)$, we find the Type ${\bf IIa}$ coins via \cref{CIIa:2} as follows
\begin{align}
	\label{eq:newCoin}
	C_{\bf IIa}=&\!\left(
	\begin{array}{cccc}
		e^{i (\phi_d - \phi_g)} \Xi c_1^2 c_2 s_2 & 
		-e^{-i \phi_e} \Xi c_1 c_2 s_1 s_3 & 
		-e^{i (\phi_h-\phi_f-\phi_g)} \Xi c_1 c_2 c_3 s_1 & 
		e^{-i\phi_f} \left(1+\Xi c_1^2 c_2^2\right) \\
		\!\!\!-e^{i (\phi_d+\phi_e+\phi_f-\phi_g-\phi_h)} \Xi c_1 c_3 s_1 s_2 \!\!\!\! & 
		e^{i (\phi_f-\phi_h)} \Xi c_3 s_1^2 s_3 & 
		e^{i (\phi_e-\phi_g)} \left(1+\Xi c_3^2 s_1^2\right) & 
		-e^{i (\phi_e-\phi_h)} \Xi  c_1 c_2 c_3 s_1\! \\
		-e^{i \phi_d} \Xi c_1 s_1 s_2 s_3 & 
		e^{i (\phi_g-\phi_e)} \left(1 + \Xi s_1^2 s_3^2\right) & 
		e^{i (\phi_h-\phi_f)} \Xi c_3 s_1^2 s_3 & 
		-e^{i (\phi_g-\phi_f)} \Xi c_1 c_2 s_1 s_3\! \\
		e^{i \phi_f} \left(1 + \Xi c_1^2 s_2^2\right) & 
		-e^{i (\phi_f+\phi_g-\phi_d-\phi_e)} \Xi c_1 s_1 s_2 s_3 & 
		\!-e^{i (\phi_h-\phi_d)} \Xi c_1 c_3 s_1 s_2 & 
		e^{i (\phi_g-\phi_d)} \Xi c_1^2 c_2 s_2 \\
	\end{array}
	\right)\!\!.
\end{align}
\end{widetext}

Now we show that \eqref{eq:newCoin} describes all possible coins when $\delta_1\in(0,\pi/2)$. It is easy to see that in this case the rank of $A$ is indeed $3$. Moreover, using our parametrization, and canceling common factors in $v_A^\dagger$ we obtain:

\begin{align}\nonumber\\[-10mm]
\ket{\psi^{ker}_A}\qquad:=\qquad\qquad\qquad c_1 s_2&\ket{L}\nonumber\\
-e^{i (\phi_d+\phi_e-\phi_g)}s_1 s_3&\ket{D}\nonumber\\
-e^{i (\phi_d+\phi_f-\phi_h)}s_1 c_3&\ket{U}\nonumber\\
+e^{i (\phi_d+\phi_f-\phi_g)}c_1 c_2&\ket{R},\label{eq:fesc}
\end{align}
a unit vector in the kernel of $A^\dagger$. We can analogously define $\ket{\psi^{ker}_B}$. Replacing an appropriate column of $A,B$ with these vectors,  one obtains \eqref{eq:newCoin} in all remaining cases involving $\delta_2\in\{0,\pi/2\}$ and / or $\delta_3\in\{0,\pi/2\}$. Therefore the formula  \eqref{eq:newCoin} covers all possible coins for rank-$3$ amplitude matrices $A$, since $\delta_1\in\{0,\pi/2\}$ implies that the rank of $A$ is $2$. (Note that \cref{eq:new} implies that $A$ has rank at least $2$, so there are no other cases remaining.)

The formula \eqref{eq:newCoin} for $C_{\bf IIa}$ may not look very intuitive, but we can describe it in a much more structured way. Let us define the one-dimensional trapping coins
\begin{align*}
C_H&=e^{-i \phi_f}\ket{R}\!\!\bra{L}+e^{i \phi_f}\ket{L}\!\!\bra{R},\\
C_V&=e^{i (\phi_e-\phi_g) }\ket{U}\!\!\bra{D}+e^{-i (\phi_e-\phi_g) }\ket{D}\!\!\bra{U},
\end{align*} 
then we get that
\begin{equation}
C_{\bf IIa}\!=\!\left(C_H\oplus  C_V\right)\left(I+\left(e^{i \eta}-1\right)\ket{\psi^{ker}_A}\!\!\bra{\psi^{ker}_A}\right).
\end{equation}
Therefore, we can view a Type {\bf IIa} coin as a modified version of a highly degenerate trapping coin which is a direct sum of one-dimensional trapping coins. In order to avoid overlaps with the Type {\bf IIb} coin class we require $\eta\neq 0$ for Type {\bf IIa} coins.

The stationary states of the coin $C_{\bf IIa}$ again come in chirally symmetric pairs which are proportional to
\begin{align}
\phantom{+} & \ket{0,0}\big(s_1 s_3 \ket{L} + c_1 s_2 e^{i\left(\phi_d+\phi_e-\phi_g\right)}\ket{D}\big) + \nonumber\\
\pm & \ket{0,1}\big(s_1 c_3 e^{i\left(\phi_h-\phi_f\right)}\ket{L} + c_1 s_2 e^{i\phi_d}\ket{U}\big) +\nonumber\\
\pm & \ket{1,0}\left(c_1 c_2 e^{i\phi_e}\ket{D} + s_1 s_3 e^{i\phi_f}\ket{R} \right) + \nonumber\\
+ & \ket{1,1}\left(c_1 c_2 e^{i\phi_g}\ket{U} + s_1 c_3 e^{i\phi_h}\ket{R}\right). \nonumber
\end{align}
In contrast to the Type {\bf I} case, the probability distribution of the above stationary states is usually non-uniform 
\begin{eqnarray}
\nonumber P(0,0) & = & \left(\cos^2\delta_1 \sin^2\delta_2 + \sin^2\delta_1 \sin^2\delta_3\right)/\, 2 , \\
\nonumber P(0,1) & = & \left(\cos^2\delta_1 \sin^2\delta_2 + \sin^2\delta_1 \cos^2\delta_3\right)/\, 2 , \\
\nonumber P(1,0) & = & \left(\cos^2\delta_1 \cos^2\delta_2 + \sin^2\delta_1 \sin^2\delta_3\right)/\, 2 , \\
\nonumber P(1,1) & = & \left(\cos^2\delta_1 \cos^2\delta_2 + \sin^2\delta_1 \cos^2\delta_3\right)/\, 2 .
\end{eqnarray}
In Section~\ref{sec:5} we identify the degenerate cases $\delta_2,\delta_3\in\{0,\frac\pi2\}$; the degeneracy again leads to two additional stationary states, which have the same form as above but the parameter $\delta_1\leftarrow\frac{\pi}{2}-\delta_1$, %$\eta\leftarrow2\pi-\eta$, 
 and some phases need to be adjusted.
E.g., when $\delta_2 = \delta_3 = 0$ the coin $C_{\bf IIa}$ becomes
\begin{align*}
\left(
\begin{array}{cccc}
0 & 0 & -e^{i (\phi_h-\phi_f-\phi_g)} \Xi c_1 s_1 & e^{-i\phi_f} \left(1+\Xi c_1^2 \right)\kern-1mm \\
0 & 0 & e^{i (\phi_e-\phi_g)} \left(1+\Xi s_1^2\right) & -e^{i (\phi_e-\phi_h)} \Xi  c_1 s_1\! \\
0 & \kern-1mm e^{i (\phi_g-\phi_e)}\kern-1mm & 0 & 0 \\
\kern-1mm e^{i \phi_f}\kern-1mm & 0 & 0 & 0
\end{array}
\right)\!\!;
\end{align*}	
the two additional stationary eigenstates have eigenvalues $\pm e^{i\frac{\eta}{2}}$, and are proportional to
\begin{align}
\mp & e^{i\left(\phi_h-\phi_f\right)}e^{i\frac{\eta}{2}}c_1\ket{0,1} \ket{L}  \pm e^{i\phi_e} e^{i\frac{\eta}{2}}s_1 \ket{1,0}\ket{D}  + \nonumber\\
+ & \ket{1,1}\left(s_1 e^{i\phi_g}\ket{U} - c_1 e^{i\phi_h}\ket{R} \right). \nonumber
\end{align}
The remaining three degenerate cases are similar.

\subsubsection{Case IIb: matrix \texorpdfstring{$A$}{A} has rank 2}

Let us now turn to the case when the matrix $A$ has rank 2, i.e., either $\delta_1=0$, implying
\begin{equation}
	a=c=f=h=0, \label{Eq:FourZeroFirst}
\end{equation}
or $\delta_1=\pi/2$, implying
\begin{equation}
	b=d=e=g=0 \label{Eq:FourZeroSecond} .
\end{equation}
We start with the case (\ref{Eq:FourZeroFirst}), when $|b|=|d|\neq 0$ (the case $|e|=|g|\neq 0$ is completely analogous). Looking at the first two columns of $A$ and $B$ in \cref{eq:Cimplicit} we get
\begin{align}
    C_{LD} &= C_{DD} = C_{RD} = C_{LU} = C_{UU} = C_{RU} = 0, \label{rank2:1}\\
	C_{UD} &= \frac{d}{b}= e^{i (\phi_g-\phi_e)}=: e^{i \gamma},\,\text{ and }\,
C_{DU} = \frac{b}{d}= e^{-i \gamma}. \label{Eq:FourZvertical}
\end{align}
The unitarity of the coin $C$ further implies that 
\begin{equation*}
C_{DL} = C_{DR} = C_{UL} = C_{UR} = 0,
\end{equation*}
i.e., the coin states describing the horizontal movement $\left\{\ket{L}, \ket{R}\right\}$ do not mix with the coin states of the vertical movement $\left\{\ket{D}, \ket{U}\right\}$. The remaining four undetermined matrix elements mixing $\ket{L}$ and $\ket{R}$ are only restricted by unitarity. Hence, they have to form a $2\times 2$ unitary matrix $C^{(1)}$, which can be parameterized, e.g., as
	\begin{equation}
C^{(1)} = \left(
		\begin{array}{rr}
		C_{LL} & C_{LR} \\
	C_{RL} & C_{RR}
		\end{array}
	\right) = e^{i\varphi}\!
	\left(
		\begin{array}{rr}
		e^{i \alpha} \cos \delta & e^{-i \beta} \sin \delta \\
		\!-e^{i \beta} \sin \delta & e^{-i \alpha} \cos \delta
		\end{array}
	\right)\!\!,
\label{rank2:C1}
	\end{equation}
with $\delta\in[0,\pi/2]$, $\varphi\in[0,\pi)$ and $\alpha, \beta, \in[0,2\pi)$. 

We conclude that the trapping coins corresponding to the case (\ref{Eq:FourZeroFirst}) must have the form
\begin{equation}	\label{Eq:CIIb1}
	C_{{\bf IIb}}^{(1)} = 
	\left(
		\begin{array}{cccc}
		\!\!\phantom{-}e^{i(\varphi + \alpha)} \cos \delta & 0 & 0 & e^{i(\varphi - \beta)} \sin \delta\\
		0 & 0 & e^{-i \gamma} & 0 \\
		0 & e^{i \gamma} & 0 & 0\\
		\!\!-e^{i(\varphi + \beta)} \sin \delta & 0 & 0 & e^{i(\varphi -\alpha)} \cos \delta
		\end{array}
	\right)\!\!.
\end{equation}

The coins corresponding to the second case (\ref{Eq:FourZeroSecond}) can be found similarly. The matrix elements can be found analogously to \cref{rank2:1,Eq:FourZvertical,rank2:C1} interchanging $L\leftrightarrow D$ and $R\leftrightarrow U$. The corresponding coins must have the \nolinebreak form
\begin{equation}\label{Eq:CIIb2}
	C_{{\bf IIb}}^{(2)} = 
	\left(
		\begin{array}{cccc}
		0 & 0 & 0 & \!\!e^{-i \phi_f}\! \\
		0 & \!\!\phantom{-}e^{i(\varphi + \alpha)} \cos \delta \, & e^{i(\varphi - \beta)} \sin \delta & \!0 \\
		0 & \!\!-e^{i(\varphi + \beta)} \sin \delta \, & e^{i(\varphi - \alpha)} \cos \delta & \!0\\
		\! e^{i \phi_f}\!\!  & 0 & 0 & \!0 
		\end{array}
	\right)\!\!.
\end{equation}

As we can see the above coins can be decomposed as the direct sum of two one-dimensional coins, and at least one of those two one-dimensional coins must be trapping. Consequently, the stationary states are also quasi one-dimensional and for the coin $C_{{\bf IIb}}^{(1)}$ have the form of
\[
\frac{\left(\ket{0,0}\ket{D} \pm e^{i\gamma}\ket{0,1}\ket{U}\right)}{\sqrt{2}}
\]
and for the coin $C_{{\bf IIb}}^{(2)}$ have the form of
\[
\frac{\left(\ket{0,0}\ket{L} \pm e^{i\phi_f}\ket{1,0}\ket{R}\right)}{\sqrt{2}}.
\]
In the degenerate case when both one-dimensional coins are trapping, then both the above ``vertical'' and ``horizontal'' stationary states appear.

Finally, note that if $\varphi=0$, then the coins $C_{{\bf IIb}}^{(1)}$ and $C_{{\bf IIb}}^{(2)}$ can be obtained from $C_{{\bf IIa}}$, for $\delta_1=0$ and $\delta_1=\pi/2$ respectively, by choosing $\eta=\pi$, and $\delta_2=\delta_3=\pi/4-\delta/2$. It is also possible to obtain instances of $C_{{\bf IIb}}^{(1)}$, $C_{{\bf IIb}}^{(2)}$ with $\varphi\neq 0$ from $C_{{\bf IIa}}$, but the range of attainable phases $\varphi$ depends on the value of $\delta$.

%%%%%%%%%%%%%%%%%%%%%%%%%%%%%%%%%%%%%%%%%%%%%%%%%%%%%%%%%%%%%%%%%%%%%%%%%%%%%%%%%%%%%%%%%%%%%%%%%%%

\section{Basic dynamical properties of the different types of trapping coins}
\label{sec:5}

In this section we investigate some basic dynamical properties of the different types of trapping coins and point out some of their characteristic differences. We focus on two things, namely the escaping initial states and the area covered by the walk. 

The escaping initial states $\ket{\psi^{ESC}}$ are those that avoid trapping. Such states have to be orthogonal to all stationary states $\ket{\psi_{st}^{(x,y)}}$. As we consider the walker starting from the origin
$$
\ket{\psi^{ESC}} = \ket{0,0}\ket{\psi_C^{ESC}}
$$
we investigate the overlap of $\ket{\psi_C^{ESC}}$ with four stationary states, namely $\ket{\psi_{st}^{(0,0)}}$, $\ket{\psi_{st}^{(0,-1)}}$, $\ket{\psi_{st}^{(-1,0)}}$ and $\ket{\psi_{st}^{(-1,-1)}}$, since the remaining ones do not overlap with $\ket{\psi^{ESC}}$, due to the $2\times 2$ support size, proven in Appendix~\ref{App:support}. We find that the coin state $\ket{\psi_C^{ESC}}$ has to be orthogonal to all local coin states $\ket{\xi^{(i,j)}}$ of the stationary states, described in (\ref{local:coin:states}). That is to say we need $\langle \psi_C^{ESC}|\xi^{(i,j)}\rangle=0$ for all $i,j\in\{0,1\}$, which is equivalent to
\begin{equation}\label{eq:esc}
\bra{\psi_C^{ESC}}A=0. 
\end{equation}
Note that \cref{eq:esc} also implies that $\ket{\psi_C^{ESC}}$ is orthogonal to the chiral counterparts of the stationary states $\ket{\psi_{st}^{(0,0)}}$, which are obtained by multiplying $\ket{\xi^{(i,j)}}$ by $(-1)^{i+j}$. Therefore $\ket{\psi_C^{ESC}}$ is indeed escaping when the above states are the only stationary eigenstates~\cite{Tate2017}. 

There can be more stationary eigenstates only if there are more than two (counted with multiplicity) constant eigenvalues of the walk operator in momentum representation~\eqref{U:kspace}. Due to chiral symmetry, the constant eigenvalues come in $\pm$ pairs,\footnote{Note that a constant eigenvalue can have non-trivial multiplicity only for coins that are direct sums of trapping one-dimensional coins, as we show in Appendix~\ref{App:support}.} therefore the number of constant eigenvalues is either $2$ or $4$ for trapping coins. When there are $4$ constant eigenvalues, the dynamics is completely trapped, and no initial state can spread further than $\pm 1$ in any directions. As we will see this only happens in degenerate cases; for Type {\bf I} coins iff $\delta_1,\delta_2\in\{0,\pi/2\}$, for Type {\bf IIa} coins iff $\delta_2,\delta_3\in \{0,\pi/2\}$ and for Type {\bf IIb} coins iff $\delta=\pi/2$.

Let us now turn to the area covered by the quantum walk. More precisely, we want to determine the set of points on the square lattice where the probability to find the walker is not negligibly (exponentially) small. This region is encompassed by the peaks in the probability distribution, which propagate in time with a constant velocity, as can be anticipated from the ballistic nature of homogeneous quantum walks. The velocities of the propagating peaks are determined by the continuous spectrum of the evolution operator $\hat U$ \cite{Konno2002,Grimmett2004,Ahlbrecht2011}. The easiest way to investigate the continuous spectrum is to employ the translational invariance of the walk and turn to the momentum representation \cite{Ambainis2001}. The Fourier transformation diagonalizes the step operator $\hat S$ and turns it into a ``point-wise'' multiplication operator given by the matrix
\begin{equation}
\label{Eq:S:k}
\widetilde{S}(k_x,k_y) = \left(\begin{array}{cccc}
	e^{-i k_x} & 0 & 0 & 0  \\
	0 & e^{-i k_y} & 0 & 0 \\
	0 & 0 & e^{i k_y} & 0 \\
	0 & 0 & 0 & e^{i k_x} 
	\end{array}\right),
\end{equation}
where $k_x$ and $k_y$ are components of the quasi-momentum\footnote{If the walk was on a finite torus with $m$ sites in both directions, then $k_x,k_y \in 2\pi\{\frac{0}{m},\frac{1}{m},\ldots,\frac{m-1}{m}\}$, the momentum-eigenstates would be $\ket{k_x,k_y}=\frac{1}{m}\sum_{x,y=0}^{m-1}e^{-i k_x x - i k_y y}\ket{x,y}$, and the step operator would be $\widetilde{S}(k_x,k_y)=\left(\bra{k_x,k_y}\otimes I\right)\hat{S}\left(\ket{k_x,k_y}\otimes I\right)$.} ranging from $-\pi$ to $\pi$.
The evolution operator in the momentum representation is block-diagonal, and it is given by the product
\begin{equation}
\widetilde U(k_x,k_y) = \widetilde{S}(k_x,k_y)\cdot C.
\label{U:kspace}
\end{equation}
In the momentum picture, the continuous spectrum of $\hat U$ is represented by the $k$-dependent eigenvalues of $\widetilde{U}(k_x,k_y)$. We show that for the Type {\bf I} and {\bf IIa} coins these eigenvalues can be written in the form 
\begin{equation}
 \lambda_\pm (k_x,k_y) = e^{i\beta} e^{\pm i\omega(k_x,k_y)} ,
 \label{cont:spec}
\end{equation}
where $\beta$ is a constant.\footnote{Note that for the Type {\bf IIb} coins the eigenvalues $\lambda_\pm$ depend only on one of the components of the quasi-momentum. We treat this case separately.} 

The rate of spreading of the quantum walk in different directions is determined \cite{Grimmett2004,Watabe2008,Ahlbrecht2011} by the properties of the function $\omega$, which can be thought of as a dispersion relation. We define the group velocities $v_x$ and $v_y$ in the $x$ and $y$ directions by
$$
v_x = \frac{\partial \omega}{\partial k_x},\quad  v_y = \frac{\partial \omega}{\partial k_y}.
$$
Asymptotically the area covered by the quantum walk corresponds \cite{Ahlbrecht2011} to the range of possible pairs $(v_x,v_y)$.  The maximal attainable group velocities can be determined by considering the Hessian matrix of $\omega$
\begin{equation}
H = \left(\begin{array}{cc}
     \frac{\partial^2 \omega}{\partial k_x^2} & \frac{\partial^2 \omega}{\partial k_x\partial k_y}  \\
     \frac{\partial^2 \omega}{\partial k_y\partial k_x} & \frac{\partial^2 \omega}{\partial k_y^2}
\end{array}\right).
\label{hess}
\end{equation}
We can find these points if we express $H$ in terms of group-velocities $v_x, v_y$, and look for points where the matrix is singular. These are the so-called caustics of the dispersion relation \cite{Ahlbrecht2011}. The set of accessible group velocities is enclosed by the points satisfying the condition $\det H = 0$; we denote its area by $\cal S$. The area covered by the quantum walk after $t$ steps is then given by ${\cal S} t^2$.

\subsection{Type {\bf I}}

In case of Type {\bf I} coins, there is a stationary eigenstate whose amplitudes form a full-rank matrix $A$ (recall $\delta_1\neq\delta_2$), thereby \cref{eq:esc} has no non-trivial solution. Hence, there is no escaping initial state. This feature was first identified in \cite{Kollar2015} and termed as strong trapping. We note that indeed the coin matrices $C_{\bf I}$ presented in (\ref{Cst}) coincide with the matrices obtained in \cite{Kollar2015}. Our analysis clarifies that strong trapping occurs if and only if the matrix $A$ has full rank.

Let us turn to the area covered by the walk. A direct calculation of the spectrum of the evolution operator in the Fourier picture $\widetilde U(k_x,k_y)$ reveals that the $k$-dependent eigenvalues can be written in the form (\ref{cont:spec}) with $\beta=0$ and the dispersion relation that reads
\begin{equation}
\omega =  -\arccos  \left[-\rho_x\cos{(k_x\!+\!\phi_x)} -\rho_y\cos{(k_y\!+\!\phi_y)} \right].
\label{omega:I}
\end{equation}
Here we have used the notation
\begin{eqnarray}
\nonumber \rho_x & = &  \cos\delta_1 \cos\delta_2, \\
 \rho_y & = & \sin\delta_1 \sin\delta_2, \\
\nonumber \phi_x & = & \phi_g -\phi_d, \\
\nonumber \phi_y & = & \phi_h -\phi_f.
\label{rho:I}
\end{eqnarray}
Note that $\omega$ becomes constant iff $\delta_1,\delta_2\in\{0,\pi/2\}$. In these degenerate cases the coin $C_{\bf I}$ is essentially a permutation matrix, so that the walker is forced to cyclically move around, and the dynamics is completely trapped.

%%%%%%%%%% Checked by Mathematica: %%%%%%%%%%%
%%%%% Continuation of test code for CI  %%%%%%
\begin{comment}
St=DiagonalMatrix[{Exp[-I kx],Exp[-I ky],Exp[I ky],Exp[I kx]}];
omega= ArcCos[ Cos[d1]Cos[d2] Cos[kx-fd+fg] + Sin[d1]Sin[d2] Cos[ky-ff+fh] ];
(St.CI - (-1) Exp[ I omega] IdentityMatrix[4]) // Det // FullSimplify[#,Assumptions -> Element[kx|ky, Reals]]&
(St.CI - (-1) Exp[-I omega] IdentityMatrix[4]) // Det // FullSimplify[#,Assumptions -> Element[kx|ky, Reals]]&
vel = D[omega /. {fd -> 0, fg -> 0, ff -> 0, fh -> 0} /. {Cos[d1] Cos[d2] -> rx, Sin[d1] Sin[d2] -> ry}, {{kx,ky}, 1}]
(H = D[vel, {{kx, ky}, 1}] // Simplify) // MatrixForm
axes1 = {Sqrt[(1 + rx^2 - ry^2 + Sqrt[(1 + rx^2 - ry^2)^2 - 4 rx^2])/2], Sqrt[(1 - rx^2 + ry^2 - Sqrt[(1 - rx^2 + ry^2)^2 - 4 ry^2])/2]}
axes2 = {rx/axes1[[1]], ry/axes1[[2]]}
Plot3D[(axes1[[1]] - axes2[[1]]) /. {rx -> Cos[d1] Cos[d2], ry -> Sin[d1] Sin[d2]}, {d1, 0, Pi/2}, {d2, 0, Pi/2}]
Plot3D[(axes1[[2]] - axes2[[2]]) /. {rx -> Cos[d1] Cos[d2], ry -> Sin[d1] Sin[d2]}, {d1, 0, Pi/2}, {d2, 0, Pi/2}]
Det[H] + rx^2 ry^2 (Cos[kx]^2 + Cos[ky]^2 + (rx^2 + ry^2 - 1) Cos[kx] Cos[ky]/(rx ry))/ (1 - (rx Cos[kx] + ry Cos[ky])^2)^2 // FullSimplify
ParametricPlot[{vel, {axes1[[1]] Sin[kx], axes1[[2]] Cos[kx]}, {axes2[[1]] Sin[ky], axes2[[2]] Cos[ky]}} /. {rx -> Cos[d1] Cos[d2], ry -> Sin[d1] Sin[d2]} /. {d1 -> Pi/7, d2 -> Pi/4}, {kx, -Pi, Pi}, {ky, -Pi, Pi}, PlotRange -> All]
\end{comment}
%%%%%%%%%%%%%%%%%%%%%%%%%%%%%%%%%%%%%%%%%%%%%%

We see that the phases $\phi_x$, $\phi_y$ do not change the overall shape of the function $\omega$, but merely shift the location of its maximum and minimum. Hence, they do not affect the range of group velocities and we can set them to zero without loss of generality, so the group velocities become
\begin{align}
\nonumber v_x & = \frac{\rho_x \sin{k_x}}{\sqrt{1-(\rho_x \cos{k_x} + \rho_y \cos{k_y})^2}}, \\
v_y & = \frac{\rho_y \sin{k_y}}{\sqrt{1-(\rho_x \cos{k_x} + \rho_y \cos{k_y})^2}}.
\label{group:vel}
\end{align}
The determinant of the Hessian matrix (\ref{hess}) in terms of the quasi-momenta $k_x$, $k_y$ is then readily obtained
\begin{align*}
\det H & = -\frac{\rho_x^2 \rho_y^2}{\left(1-(\rho_x \cos{k_x} + \rho_y \cos{k_y})^2\right)^2} \times \\
& \times\left(\cos^2{k_x} + \cos^2{k_y} + \frac{\rho_x^2+\rho_y^2 - 1}{\rho_x \rho_y}\cos{k_x}\cos{k_y}\right)\!.
\end{align*}
Note that it depends only on cosines of the quasi-momenta. To express $\det H$ in terms of the group velocities, we take the squares of the equations (\ref{group:vel}) and determine $\cos{k_x}$ and $\cos{k_y}$ as functions of $v_x$, $v_y$. The resulting expression for $\det H$ is rather convoluted, but one can show that it vanishes for $v_x$, $v_y$ lying on two ellipses 
\begin{equation}
{\cal E}_i : \frac{v_x^2}{a_i^2} + \frac{v_y^2}{b_i^2} = 1,\quad i=1,2,
\label{ellipse}
\end{equation}
\begin{figure}[ht!] 
	\begin{center}
		\vspace{-0mm}
		\includegraphics[width=0.5\textwidth]{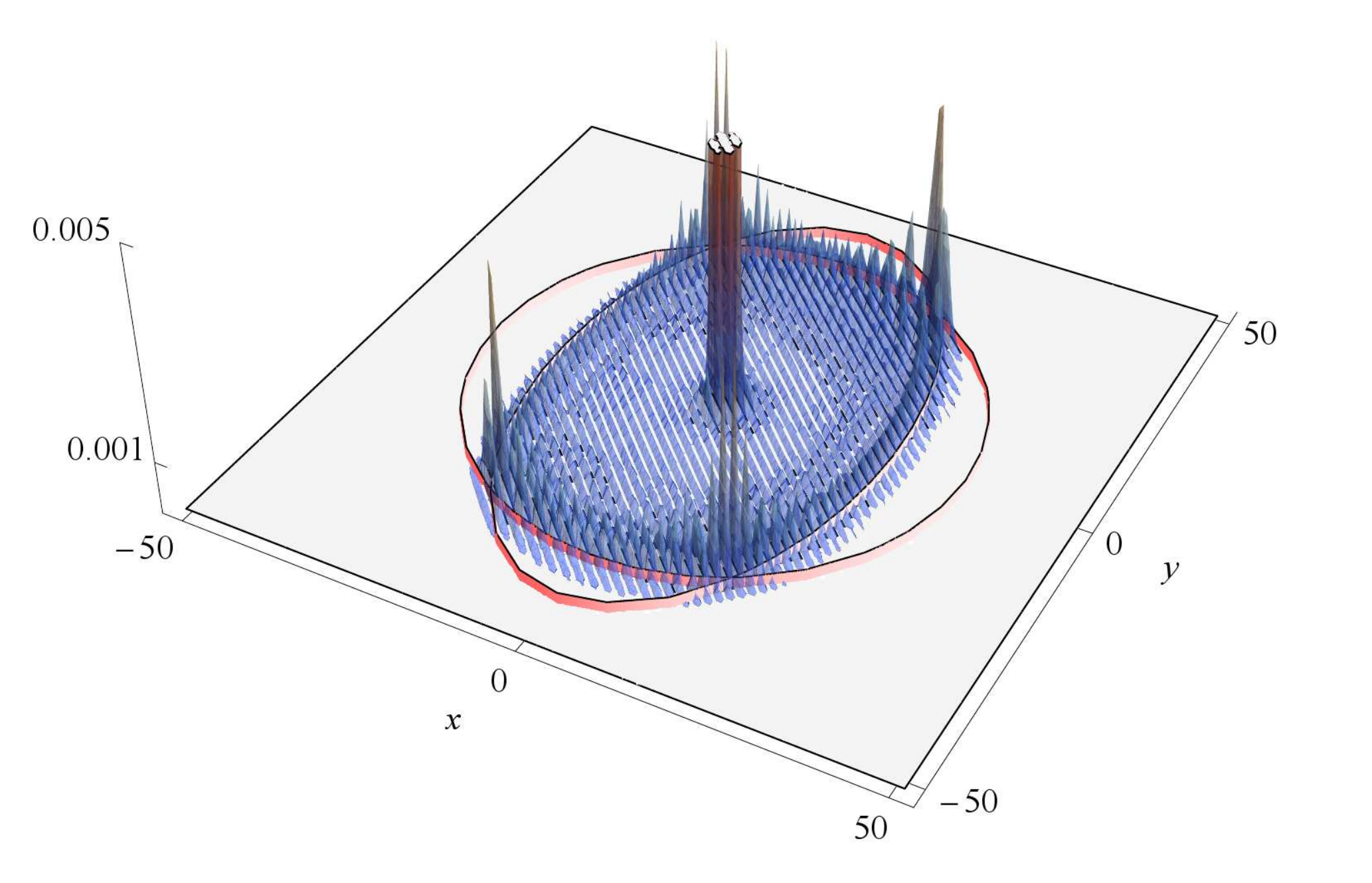}
		\vspace{-7mm}
	\end{center}
	\caption{Probability distribution after 50 steps of the quantum walk with the coin $C_{\bf I}$ and the parameters $\delta_1 = \frac{\pi}{3}$, $\delta_2 = \frac{\pi}{4}$. All phases $\phi_j$ were set to zero. For this choice of parameters the ellipse ${\cal E}_1$ becomes a circle with diameter $a_1 = b_1 = 1/\sqrt{2}$, and the second ellipse ${\cal E}_2$ has semi-axes $a_2 = \frac{1}{2}$ and $b_2 = \sqrt{3}/2$. The red curves correspond to the re-scaled ellipses ${\cal E}_i$ where we replace $v_x, v_y$ by $\frac{x}{t},\frac{y}{t}$. The initial coin state was chosen as $\ket{\psi_C} = \frac{1}{2}\left(\ket{L} + i\ket{D} + i\ket{U} + \ket{R}\right)$ resulting in a symmetric probability distribution. Only points with probability greater than $10^{-5}$ are plotted; the covered area accurately fits the intersection of the interiors of the two ellipses.} 
	\label{fig:rank4}
\end{figure}
The semi-axes of the first ellipse are given by
\begin{align}
\nonumber a_1 & = \sqrt{\frac{1+\rho_x^2 - \rho_y^2 + \sqrt{(1+\rho_x^2-\rho_y^2)^2-4\rho_x^2}}{2}},\\
b_1 & = \sqrt{\frac{1-\rho_x^2 + \rho_y^2 - \sqrt{(1-\rho_x^2+\rho_y^2)^2-4\rho_y^2}}{2}},
\label{semiaxes1:CI}
\end{align}
while for the second ellipse they read 
\begin{equation}
a_2 = \frac{\rho_x}{a_1}, \qquad
b_2 = \frac{\rho_y}{b_1}.
\label{semiaxes2:CI}
\end{equation}

Let us denote by ${\cal E}_i^{\circ}$ the interior points of the ellipse ${\cal E}_i$. For the points that are interior of one ellipse but exterior of the other the transformation $(k_x,k_y)\rightarrow(v_x,v_y)$ is not defined. We conclude that the range of accessible group velocities for the quantum walk with the Type {\bf I} coin is given by ${\cal E}_1^\circ\cap{\cal E}_2^\circ$. We note that the ellipses cannot coincide since for Type {\bf I} coins we require $\delta_1\neq \delta_2$. For illustration, in Figure~\ref{fig:rank4} we show the probability distribution of the quantum walk with a Type {\bf I}. 

\begin{figure}[ht!] 
	\begin{center}
	\hskip-8mm
	\begin{tikzpicture}
	\node[inner sep=0pt] at (0,0) (areas)
	{\includegraphics[width=.4\textwidth]{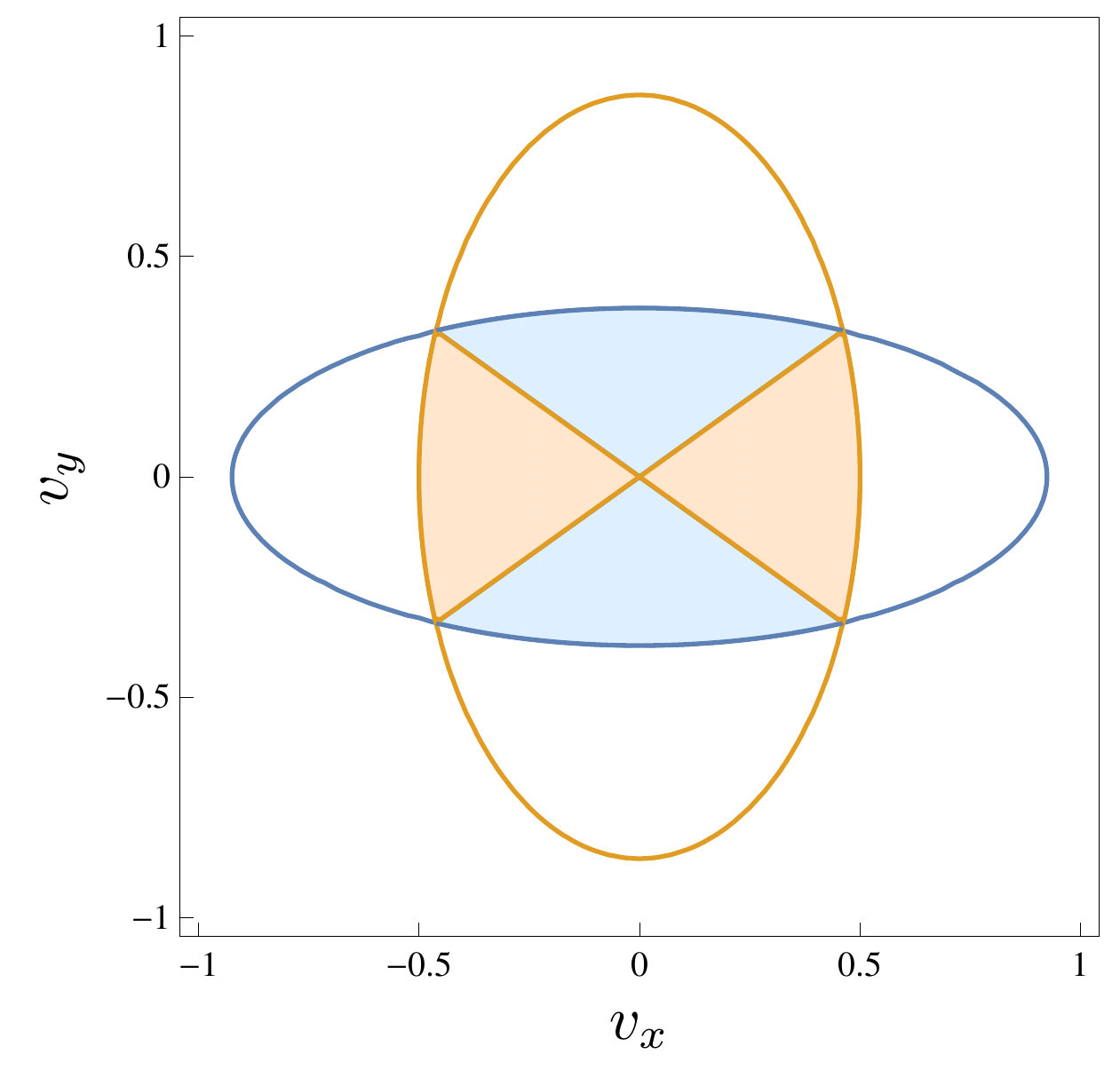}};
	\node[] at (0.525, 1)	(t1) {$\theta_1$};
	\node[] at (1.5, 0.33)		(t2) {$\theta_2$};	
	\end{tikzpicture}
	\vspace{-3mm}
	\end{center}
\caption{The ellipses ${\cal E}_1$ (blue curve) and ${\cal E}_2$ (orange curve) for the parameters $\delta_1 = \frac{\pi}{3}$ and $\delta_2 = \frac{\pi}{8}$. The area of their overlap can be decomposed into two ellipse sectors of ${\cal E}_1$ (blue regions) and two ellipse sectors of ${\cal E}_2$ (orange regions).}
\label{fig:area}
\end{figure}

Let us now determine the area ${\cal S}$ of the set of attainable group velocities, which is given by the overlap of the two centered ellipses ${\cal E}_1$ and ${\cal E}_2$. We can decompose it into two ellipse sectors of ${\cal E}_1$ (with the same area ${\cal S}_1$) and two ellipse sectors of ${\cal E}_2$ (with the same area ${\cal S}_2$), see Figure~\ref{fig:area}. The area of the overlap is then given by
\begin{align}\label{overlap}
{\cal S} & = 2 {\cal S}_1 + 2 {\cal S}_2  = \theta_1 a_1 b_1 + \theta_2 a_2 b_2, 
\end{align}
where $\theta_i$ are the parametric angles defined by the four intersection points $(\pm v_x^{int}, \pm v_y^{int})$ of the two ellipses, i.e. 
\begin{align*}
\nonumber \theta_1  = 2\arcsin \left(\frac{v_x^{int}}{a_1} \right),\,\, 
\theta_2  = 2\arccos \left(\frac{v_x^{int}}{a_2} \right).
\end{align*}
The first coordinate $\pm v_x^{int}$ of the intersection points can be computed from the length of the semi-axes as follows 
\begin{equation*}
v_x^{int} = \sqrt{\left\lvert \frac{b^2_1 - b^2_2}{a^2_2 b^2_1 -  a^2_1 b^2_2 } \right\rvert} a_1 a_2 .
\end{equation*}

\begin{figure}[ht!] 
	\begin{center}
	\vspace{-0mm}
	\includegraphics[width=0.4\textwidth]{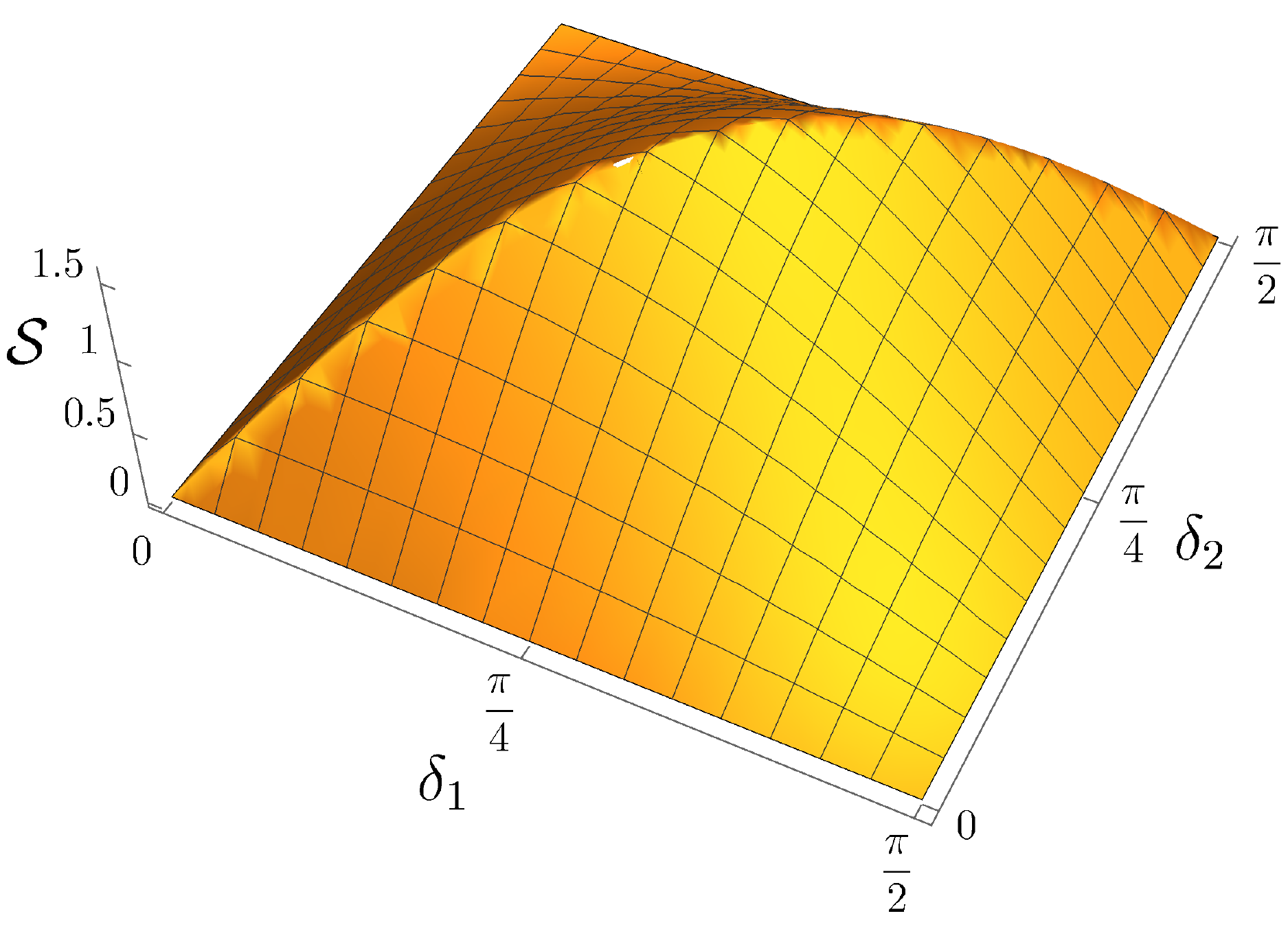}
	\vspace{-3mm}	
	\end{center}
\caption{Area $\cal S$ of the set of attainable group velocities for the walk with the Type $\bf I$ coin (\ref{overlap}) as a function of the coin parameters $\delta_1$ and $\delta_2$. %Note that the diagonal should be excluded.
}
\label{fig:area:3D}
\end{figure}

For illustration we show in Figure \ref{fig:area:3D} the area ${\cal S}$ as a function of the coin parameters $\delta_1$ and $\delta_2$. The covered area changes significantly for different pairs of $\delta_1$ and $\delta_2$. In case $\delta_1\in\{0,\pi/2\}$ or $\delta_2\in\{0,\pi/2\}$ the walker does not spread in one of the directions and the dynamics is essentially one dimensional, so the covered area is very small. In the other extreme when $\delta_1 \approx \delta_2 \approx \pi /4$ (remember we excluded $\delta_1 = \delta_2 $) the ellipses almost become two identical circles, maximizing the covered area.

\subsection{Type {\bf IIa}}

In this case the rank of $A$ is $3$, therefore \cref{eq:esc} has a unique solution, which is the state described in \cref{eq:fesc}, i.e. $\ket{\psi_C^{ESC}} = \ket{\psi_A^{ker}}$. 

Let us now investigate the continuous spectrum \eqref{cont:spec}. A direct computation shows that for Type {\bf IIa} coins we can set $\beta = \frac{\eta-\pi}{2}$ and the function $\omega$ has the same structure as \cref{omega:I}. The phases $\phi_x$, $\phi_y$ remain the same as for the Type {\bf I} coins, while the parameters $\rho_x$ and $\rho_y$ are given by
\begin{align}
\nonumber \rho_x & = \cos^2{\delta_1} \sin{2\delta_2} \sin{\frac{\eta}{2}}, \\
 \rho_y & = \sin^2{\delta_1} \sin{2\delta_3} \sin{\frac{\eta}{2}}.
\label{rho:II}
\end{align}
Similarly to the previous case $\omega$ becomes constant iff $\delta_2,\delta_3\in\{0,\pi/2\}$ (remember that $\delta_1\in(0,\pi/2)$ and $\eta\neq 0$). These degenerate cases result in a completely trapped dynamics, but interestingly the corresponding coin matrices do not have permutation structure.

\begin{figure}[ht!] 
	\begin{center}
\includegraphics[width=0.5\textwidth]{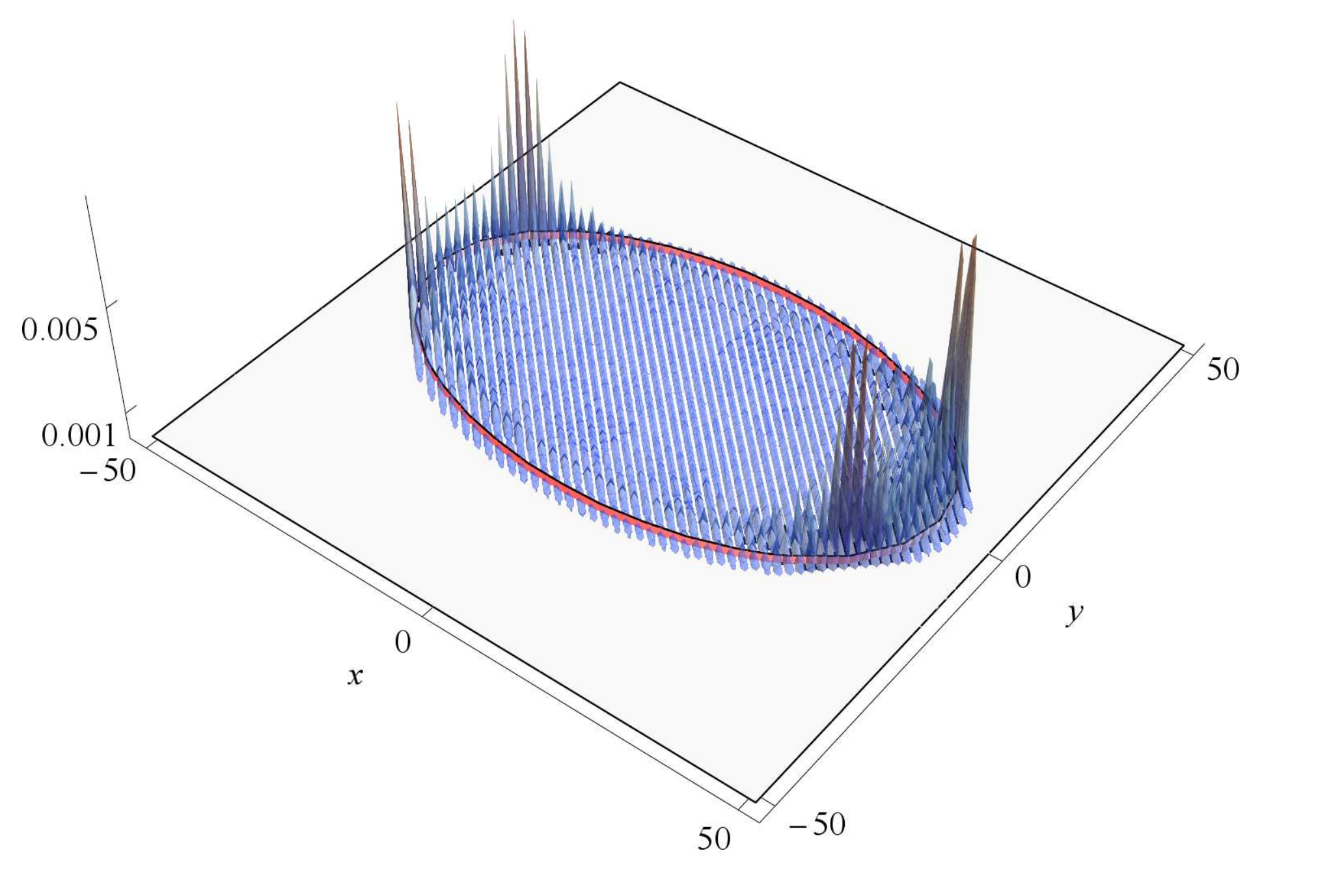}
	\end{center}
    	\caption{Probability distribution after 50 steps of the quantum walk with the coin $C_{\bf IIa}$ and the parameters $\delta_1 = \frac{\pi}{6}$, $\delta_2 = \delta_3 = \frac{\pi}{4}$ and $\eta = \pi$. All phases $\phi_j$ were set to zero. For this choice the range of attainable group velocities is given by a single ellipse ${\cal E}_1 = {\cal E}_2$ with semi-axes $a_1 = \frac{\sqrt{3}}{2}$ and $b_1 = \frac{1}{2}$. The re-scaled ellipse is plotted with the red curve. The initial coin state was chosen as the escaping state (\ref{eq:fesc}) which results in a symmetric probability distribution without the central trapping peak. Only points with probability greater than $10^{-5}$ are plotted. } 
\label{fig:rank3}
\end{figure}

Since $\omega$ has the same form as for the Type ${\bf I}$ coin we use the previously derived results and find that the area covered by the walk is again determined by the intersection of ${\cal E}_1^\circ$ and ${\cal E}_2^\circ$, see Figure~\ref{fig:rank3}. Unlike for the Type {\bf I} solutions, the ellipses can coincide. Indeed, for $\eta=\pi$ and $\delta_2=\delta_3=\frac{\pi}{4}$ we find that the semi-axes of ${\cal E}_1$ and ${\cal E}_2$ are the same and read
\begin{align}
\nonumber a_1 & = a_2 = \cos{\delta_1}, \\
 b_1 & = b_2 = \sin{\delta_1}.
\label{x1=x2}
\end{align}
We note that in this case the matrix $C_{\bf IIa}$ coincides (up to a permutation due to a different ordering of the basis states of the coin space) with the coin considered in \cite{Watabe2008}, where $p=\cos^2{\delta_1}$ and $q=1-p = \sin^2{\delta_1}$. Additionally, choosing $\delta_1=\frac{\pi}{4}$ the matrix $C_{\bf IIa}$ reduces to the $4 \times 4$ Grover coin explored in detail in \cite{Inui2004}. For this particular coin the range of attainable group velocities is given by a circle of radius $\frac{1}{\sqrt{2}}$.

\subsection{Type IIb}

In this case the rank of $A$ is $2$, therefore \cref{eq:esc} has multiple solutions and the escaping states form a two-dimensional subspace, unless $\delta=\pi/2$ and the coin is a direct sum of trapping one-dimensional coins. For less degenerate coins $C^{(1)}_{\bf IIb}$, every ``horizontal'' state is escaping
\begin{equation*}
\ket{\psi_C^{ESC}} = \psi_L \ket{L} + \psi_R \ket{R},
\end{equation*}
and for $C^{(2)}_{\bf IIb}$ coins, every ``vertical'' state is escaping
\begin{equation*}
\ket{\psi_C^{ESC}} = \psi_U \ket{U} + \psi_D \ket{D}.
\end{equation*}

Let us turn to the spreading properties of the walks. It can be anticipated from the form of the matrices (\ref{Eq:CIIb1}) and (\ref{Eq:CIIb2})  that the walks are essentially one-dimensional. Indeed, for the $C^{(1)}_{\bf IIb}$ the continuous spectrum of the evolution operator $\hat U$ is given by
$$
\lambda_\pm = e^{i \varphi}e^{\pm i\omega(k_x)},
$$
where the function $\omega(k_x)$ is
$$
\omega(k_x) = -\arccos{(\cos{\delta}\cos{(k_x-\alpha)})}.
$$
Since $\omega$ is independent of $k_y$ the group velocity in the $y$ direction vanishes, i.e. the walk spreads only in the $x$ direction with the group velocity 
$$
v_x = \frac{\cos \delta \sin (\alpha-k_x)}{\sqrt{1-\cos ^2(\delta) \cos
		^2(\alpha-k_x)}}.
$$

The coin parameter $\delta$ determines the rate of spreading in the $x$ direction, as the maximum of the group velocity $v_x$ is given by
$$
\max{v_x} = \cos{\delta}.
$$
Hence, after $t$ steps the two propagating peaks in the probability distribution are located approximately at positions $\pm t\cos{\delta }$.

\begin{figure}[ht!] 
	\begin{center}
\includegraphics[width=0.475\textwidth,trim=0 0 0 30mm, clip]{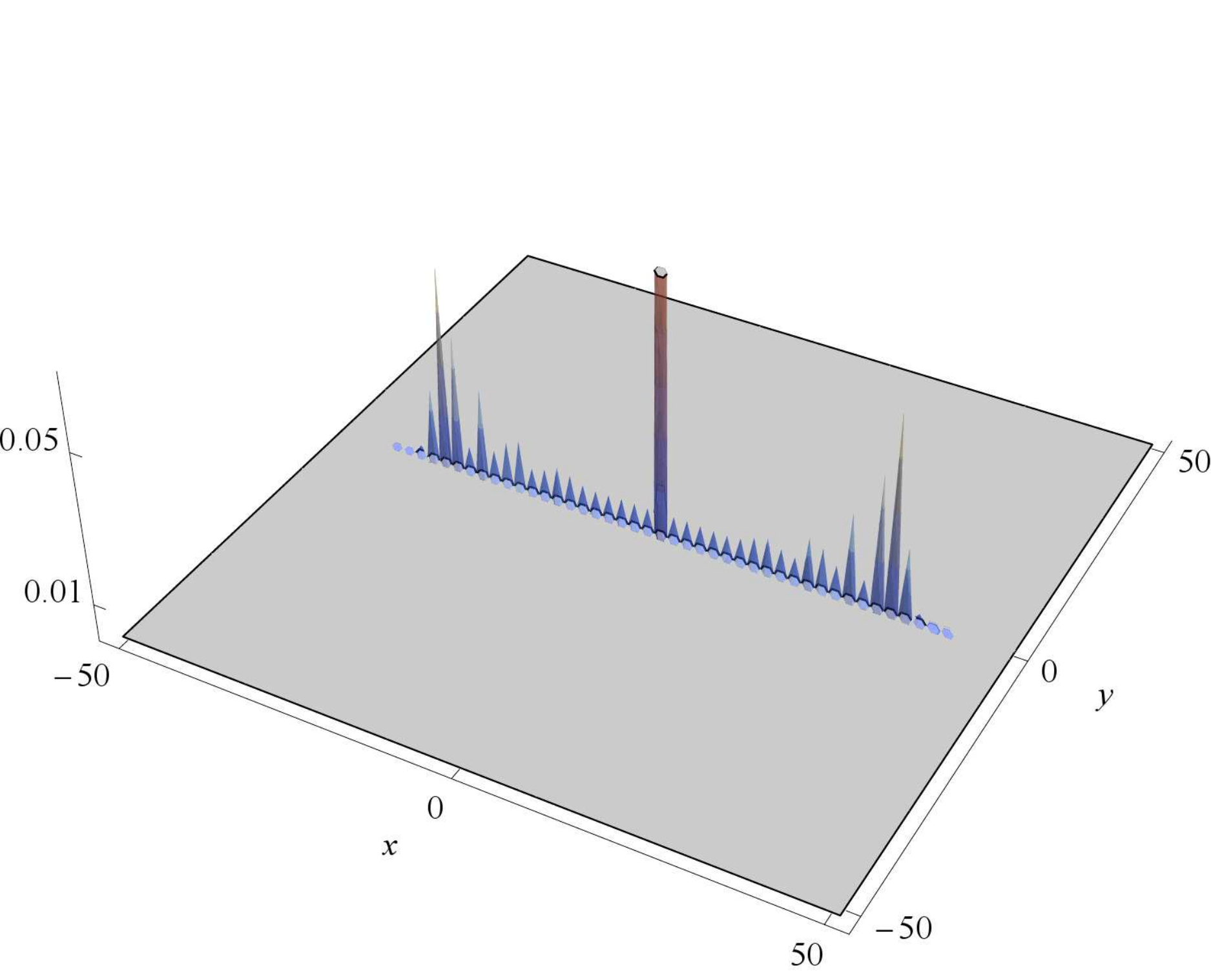}
\vskip-10mm
	\end{center}
    	\caption{Probability distribution of the quantum walk with the coin $C^{(1)}_{\bf IIb}$ after 50 steps of the walk. The coin parameters were chosen according to $\varphi = \alpha = \beta = \gamma = 0$ and $\delta = \frac{\pi}{4}$. The initial state was chosen as $\ket{\psi_C} = \frac{1}{2}\left(\ket{L} + \ket{D} + \ket{U} + i\ket{R}\right)$ resulting in a symmetric probability distribution. The central peak corresponds to the trapping effect. Clearly, the walker spreads only in the $x$ direction.} 
\label{fig:rank2}
\end{figure}

Figure~\ref{fig:rank2} illustrates the probability distribution of a quantum walk with the coin $C^{(1)}_{\bf IIb}$. The coin $C^{(2)}_{\bf IIb}$ leads to similar behavior but with spreading in the $y$ direction. 

\section{Summary}
\label{sec:6}

In this paper we have studied four-state discrete-time quantum walks on a two-dimensional lattice, where the quantum particle is allowed to move horizontally to the left and right and vertically up and down. We classified all such quantum walk operators that exhibit trapping, manifested in a non-vanishing peak of the position probability distribution situated at the initial position of the walker. This effect is linked to the existence of point spectrum of the evolution operator, which we ``shifted'' to eigenvalues $\pm 1$, due to the irrelevance of global phases.

We explicitly constructed all trapping coin operators, using the observation that the stationary eigenstate can be confined to a $2\times 2$ patch of the lattice.
Three distinct types of parameterized solutions were found, which we explicitly describe (up-to a global phase factor). The first type of coins $C_{\mathbf I}$ have seven real parameters. This family exhibits strong trapping, i.e., any walk started from a single site have a non-vanishing component trapped at the initial site.
The second type of coins $C_\mathbf{IIa}$ have nine real parameters, and they do not exhibit the stronger version of trapping, except for a degenerate case. In the non-degenerate case there is always a unique escaping state for which the probability of staying at the initial site vanishes over time. For instance, the well-known Grover coin is within this coin class.
Finally, the third type of coins $C_\mathbf{IIb}$ are quasi one-dimensional since they can be written as a direct sum of one-dimensional coins, at least one of which must be trapping.

We have also determined the area covered by the spreading component for each types of walk. For the first class of coins, the area covered by the wave function of the walker can be well estimated by the cross section of two different ellipses. The situation is similar for the second class of coins, except the ellipses can in certain cases merge to a single one. For the last type of coins the walk is characterized by quasi-one-dimensional dynamics either along the horizontal or vertical direction.

Understanding the spreading properties of quantum walks may be useful in situations where one would like to manipulate or shape the transport properties of media modeled by homogeneous quantum walks. For example switching between different types of coins, might turn off trapping and recover ballistic spreading. 

In summary, we provided a full classification for the basic quantum walk on the two-dimensional square lattice by analyzing the stationary eigenstates. A similar, constructive approach might be applicable for higher dimensional lattices and possibly to other regular graphs as well. However, we note that for non-square lattices one needs to be careful about the choice of the shift operator. For example, there are no trapping coins on a triangular lattice with a moving shift operator \cite{kollar2010}, whereas for the flip-flop (or reflecting) shift operator the quantum walk with the Grover coin has stationary states \cite{mares2019}. It would be interesting to see whether our techniques can be applied to characterize trapping coins in the latter case.

\begin{acknowledgments}

I.~J.~and M.~\v S.~received support from the Czech Grant Agency (GA\v CR) under grant No.~17-00844S and from MSMT RVO 14000. This publication was funded by the project "Centre for Advanced Applied Sciences", Registry No.~CZ.$02.1.01/0.0/0.0/16\_019/0000778$, supported by the Operational Programme Research, Development and Education, co-financed by the European Structural and Investment Funds and the state budget of the Czech Republic. T.~K.~was supported by the National Research, Development and Innovation Office of Hungary (Project Nos.~K124351 and 2017-1.2.1-NKP-2017-00001). A.~G.~was supported by ERC Consolidator Grant QPROGRESS and partially supported by QuantERA project QuantAlgo 680-91-034, and also by the Institute for Quantum Information and Matter, an NSF Physics Frontiers Center (NSF Grant PHY-1733907). I.~T.~would like to acknowledge the financial support from Technical University of Ostrava under project no.~SP2018/44.
\end{acknowledgments}

%%%%%%%%%%%%%%%%%%%%%%%%%%
\appendix

\section
{\texorpdfstring{$\pmb{2\times 2}$}{\bf 2 x 2} support of eigenstates}
\label{App:support}

Our construction of trapping coins is based on the properties of stationary eigenstates. Namely, we use that there must exist a localized eigenstate with a support of size at most $2 \times 2$, i.e., it has the form given in \cref{Eq:StStatexy}. 

In order to prove this statement we turn to the momentum picture, where the evolution operator $\hat U$ is represented by the matrix $\widetilde U(k_x,k_y)$ given by (\ref{U:kspace}). Due to~\cite{Tate2017} we know that if $1$ is an infinitely degenerate eigenvalue of $\hat{U}$, it is also an eigenvalue of $\widetilde U(k_x,k_y)$ for every $k_x,k_y\in [-\pi,\pi]$, and so $\det(\widetilde U(k_x,k_y) - I)=0$. Our goal is to find a parameterized eigenvector with bounded Fourier spectrum, i.e., vectors $\xi^{(j,\ell)}$ for $j,\ell\in \{0,1\}$ such that for all $k_x,k_y\in [-\pi,\pi]$ we have
\begin{equation}
\label{app:psi}
(\widetilde U(k_x,k_y) - I) \underbrace{\sum_{j,\ell=0}^{1}e^{i j k_x}e^{i \ell k_y}\xi^{(j,\ell)}}_{\tilde{\psi}(k_x,k_y):=}=0.
\end{equation}
%%%%%%%%%%%%%%%%%%%%%%%%%%%%%%%%%%%%%%%%%%%%%%%%%%%%%%
%%%%% Checking the sign in the Fourier transform %%%%%
%%%%%%%%%%%%%%%%%%%%%%%%%%%%%%%%%%%%%%%%%%%%%%%%%%%%%%
\begin{comment}
\begin{align*}
\bra{k,L}\tilde{S}\ket{k,L} = e^{-ik}  & \Leftrightarrow \tilde{S}=\mathrm{QFT} \circ S \circ \mathrm{QFT}^{-1} \\
&\Leftrightarrow \tilde{U}=\mathrm{QFT} \circ U \circ \mathrm{QFT}^{-1} \Rightarrow \\
\tilde{U}\ket{\tilde{\psi}} = \ket{\tilde{\psi}} &\Leftrightarrow U\left( \mathrm{QFT}^{-1}\ket{\tilde{\psi}}\right) =  \mathrm{QFT}^{-1}\ket{\tilde{\psi}}
\end{align*} 
\end{comment}
%%%%%%%%%%%%%%%%%%%%%%%%%%%%%%%%%%%%%%%%%%%%%%%%%%%%%%
By applying the inverse Fourier transform to $\tilde{\psi}$ one can see that it corresponds to an eigenvalue $1$ eigenstate of \nolinebreak $\hat U$ localized at the vertices $(0,0)$, $(0,1)$, $(1,0)$ and $(1,1)$:
\begin{equation*}
\ket{\psi_{st}^{(0,0)}}=\sum_{j,\ell=0}^{1}\ket{j,\ell}\ket{\xi^{(j,\ell)}}.
\end{equation*}
Let us introduce the notation $x:=e^{i k_x}$, $y:=e^{i k_y}$ so that 
$$
\widetilde{S}=\mathrm{diag}(1/x,1/y,y,x).
$$
We will think about $\widetilde U=\widetilde S C$ as a matrix with Laurent polynomial entries in $x,y$. 

A Laurent polynomial $f$ in variable $x$ of degree at most $n$ over the ring $R$ is an expression of the form 
$$
f(x)=\sum_{i=-n}^{n}c_i x^i,
$$
for some $c_i\in R$ coefficients. We denote the set of Laurent polynomials in variable $x$ by $R[x^{\pm 1}]$, and define 
\begin{align*}
\max\deg(f(x)) & :=\max\{i\colon c_i\neq 0\} , \\
\min\deg(f(x)) & :=\min\{i\colon c_i\neq 0\} .
\end{align*}
We will use the handful property~\cite[Ch.~VII\S 3.4]{Bourbaki1998commutative} that if $R[x]$ is a unique factorization domain (UFD), then so is $R[x^{\pm 1}]$. In particular, two-variate complex Laurent polynomials, denoted by $\mathbb{C}[x^{\pm 1},y^{\pm 1}]$, form a UFD. We will also use the fact that if $f\in \mathbb{C}[x^{\pm 1},y^{\pm 1}]$ is zero for all $(x,y) \in S\times S'$ for some infinite sets $S,S'\subseteq\mathbb{C}$, then $f\equiv 0$, which directly follows from the analogous statement for polynomials~\cite{Alon1999}.\footnote{We denote the equality of (vectors consisting of) Laurent polynomials by $\equiv$ to improve readability.} For brevity in the rest of this appendix when we say that a Laurent polynomial has degree-$n$ we mean that it has degree at most~$n$.

We already know that $\det(\widetilde U - I)= 0$ for every $x,y\in \mathbb{C}$ of unit modulus, which then implies $\det(\widetilde U - I)\equiv 0$, and 
\begin{equation}\label{eq:detTransfer}
0\equiv\det(\widetilde S^{-1})\det(\widetilde U - I)\equiv \det(\underbrace{C- \widetilde S^{-1}}_{D:=}),
\end{equation}
\vskip-4mm\noindent
where
\begin{equation}\label{eq:det}
D= \left(\begin{array}{cccc}
C_{11}-x & C_{12} & C_{13} & C_{14} \\
C_{21} & C_{22}-y & C_{23} & C_{24} \\
C_{31} & C_{32} & C_{33}-\frac{1}{y} & C_{34} \\
C_{41} & C_{42} & C_{43} & C_{44}-\frac{1}{x}
\end{array}\right).
\end{equation}
In order to satisfy (\ref{app:psi}) it suffices to find a vector $\tilde{\psi}\in(\mathbb{C}[x^{\pm 1},y^{\pm 1}])^4$ with entries that are ordinary polynomials of degree 1 in each variable, such that $D\tilde{\psi}\equiv 0$, since then 
$$
(\widetilde U - I)\tilde{\psi}\equiv \widetilde S D \tilde{\psi}\equiv \widetilde S 0 \equiv 0.
$$

We proceed by case separation. First we treat the case when all matrix minors $M$ of $D$ with size $3\times 3$ have $\det(M) \not\equiv 0$. Let $M$ be such a matrix minor that we get by deleting row and column $i$. We can formally compute the inverse matrix $M^{-1}$ using Cramer's rule, where each matrix element is a subdeterminant divided by $\pm\det(M)$. We take the matrix 
$$
M':= \det(M)M^{-1},
$$
which is a matrix with Laurent polynomial entries such that 
$$
M M'=M' M=\det(M) I.
$$
Let $\bar{v}^{(i)}$ be the 3-dimensional vector that we get from the $i$-th column of $D$ by deleting its $i$-th entry, and let 
$$
\bar{w}^{(i)}:=M' \bar{v}^{(i)}.
$$
Finally let $w^{(i)}$ be the 4-dimensional vector that we get by inserting an $i$-th entry with value $-\det(M)$. We claim that $D w^{(i)}\equiv 0$, while $w^{(i)}\not\equiv 0$; the latter follows from $w^{(i)}_i\equiv -\det(M)\not \equiv 0$.

For all but the $i$-th coordinate of $D w^{(i)}$ we immediately get by construction that its value is $\equiv 0$. Now we prove that the $i$-th coordinate is zero as well. Let $D'$ be the matrix we get from $D$ by multiplying the $i$-th column by $\det(M)$, and let $D''$ the matrix we get from $D'$ by adding $D w^{(i)}$ to its $i$-th column. Now observe that 
$$
\det(D')\equiv \det(M)\det(D)\equiv 0,
$$
and that $\det(D'')\equiv 0$, because its $i$-th column is a linear combination of its other columns. Therefore 
\begin{equation}\label{eq:detOfinally}
0\equiv \det(D'')-\det(D')\equiv(D w^{(i)})_i \det(M),
\end{equation}
where the last equality holds because the $(i,i)$ matrix element of $D''-D'$ is $(D w^{(i)})_i$ and the rest of $D''-D'$ is zero, moreover the corresponding matrix minor of both $D''$ and $D'$ is $M$. Since $\det(M) \not\equiv 0$, \cref{eq:detOfinally} implies that $(D w^{(i)})_i\equiv 0$.

Considering $i=4$ we observe that each coordinate of $w^{(4)}$ is a complex linear combination of (sub)determinants of $M$, implying 
\begin{align}
 0&\leq \min\deg(w^{(4)}(x))\leq \max\deg(w^{(4)}(x))\leq 1, \label{eq:mmdeg1}\\
-1&\leq \min\deg(w^{(4)}(y))\leq \max\deg(w^{(4)}(y))\leq 1.\label{eq:mmdeg2}
\end{align}
By symmetry we also get 
\begin{align}
 0&\leq \min\deg(w^{(3)}(y))\leq \max\deg(w^{(3)}(y))\leq 1, \label{eq:mmdeg3}\\
-1&\leq \min\deg(w^{(3)}(x))\leq \max\deg(w^{(3)}(x))\leq 1.\label{eq:mmdeg4}
\end{align}

If $(x,y)$ is such that $\det(M(x,y))\neq 0$, then the kernel of $D(x,y)$ is one-dimensional; consequently $w^{(3)}(x,y)$ and $w^{(4)}(x,y)$ are linearly dependent. This implies that for all $x,y \in \mathbb{C}$ and $j,\ell\in\{1,\ldots,4\}$ 
\begin{equation*}
\det(M)w^{(3)}_j w^{(4)}_\ell=\det(M)w^{(3)}_\ell w^{(4)}_j,
\end{equation*}  
and since $\det(M)\not \equiv 0$ it also implies
\begin{equation}\label{eq:ratioPoly}
w^{(3)}_j w^{(4)}_\ell\equiv w^{(3)}_\ell w^{(4)}_j.
\end{equation} 
As $w^{(i)}_i\not \equiv  0$, applying \cref{eq:ratioPoly} with $j=3$ and $\ell=4$ implies that neither $w^{(3)}_4\equiv 0$ nor $w^{(4)}_3\equiv 0$. 

Since $\mathbb{C}[x^{\pm 1},y^{\pm 1}]$ is UFD, we can write $w^{(3)}_4/w^{(4)}_4$ in lowest terms $f/g$
where the Laurent polynomials $f$ and $g$ have no non-trivial common factors. We can also assume without loss of generality that $\min\deg(g(x))=\min\deg(f(y))=0$ (otherwise we can take $g\cdot m$ and $f\cdot m$ for the appropriate monomial $m=x^k y^n$).
Then applying \cref{eq:ratioPoly} with $\ell=4$ we get $w^{(3)}_j g \equiv f w^{(4)}_j$ so using the UFD property we get that $w^{(3)}_j/f\equiv w^{(4)}_j/g$ are Laurent polynomials for all $j\in\{1,\ldots, 4\}$. Since for $a,b\in \mathbb{C}[x^{\pm 1},y^{\pm 1}]$ we have
\begin{align*}
\min\deg((ab)(x))&=\min\deg(a(x))\min\deg(b(x)),\\
\max\deg((ab)(x))&=\max\deg(a(x))\max\deg(b(x)),
\end{align*}
the assertion 
$$
\min\deg(g(x))=\min\deg(f(y))=0,
$$ together with \cref{eq:mmdeg1,eq:mmdeg2,eq:mmdeg3,eq:mmdeg4} imply the desired property
\begin{align*}
0&\leq \min\deg((w^{(4)}/g)(x))\leq \max\deg((w^{(4)}/g)(x))\leq 1, \\
0&\leq \min\deg((w^{(3)}/f)(y))\leq \max\deg((w^{(3)}/f)(y))\leq 1.
\end{align*}
Let 
\begin{align*}\\[-10mm]
\tilde{\psi}:=w^{(3)}/f\equiv w^{(4)}/g\in (\mathbb{C}[x^{\pm 1},y^{\pm 1}])^4 . 
\end{align*}
We see that it is and ordinary polynomial of degree 1 in each variable, which satisfies
$$
0\equiv D w^{(4)}\equiv g D \tilde{\psi},
$$
i.e., $0\equiv D \tilde{\psi}$. Hence, $\tilde{\psi}$ is the desired eigenvector of $\widetilde{U}$.

It remains to check the case when there is an $i\in\{1,2,3,4\}$ such that the matrix minor $M$ that we obtain by deleting row and column $i$ of $D$ has $\det(M) \equiv 0$. Suppose that $i=4$, then the coefficient of $x$ in $\det(M)$ equals the determinant of 

\begin{align*}\\[-10mm]
\widetilde M = 	\left(
\begin{array}{ll}
C_{22}-y & C_{23} \\
C_{32} & C_{33} -\frac{1}{y}
\end{array}\right),
\end{align*}
the middle $2\times 2$ matrix of $D$, implying $\det(\widetilde M)\equiv 0$. The coefficients of $y$ and $1/y$ in $\det(\widetilde M)$ come from the diagonal elements ${C}_{22},{C}_{33}$, which then must be zero. Thus the constant term in $\det(\widetilde M)$ equals $1-{C}_{23}{C}_{32}$. Since $|{C}_{23}|\leq 1$ and $|{C}_{32}|\leq 1$ we must have $|{C}_{23}|=|{C}_{32}|=1$ and ${C}_{23}={C}_{32}^*$. This implies that all matrix elements in the second and third columns of $C$ equal $0$ except ${C}_{23}$ and ${C}_{32}$, and so the vector 
$$
\tilde{\psi}:=(0,1,{C}_{32}y,0)^T ,
$$ 
is in the kernel of $D$. The proof for the other values $i\in\{1,2,3\}$ follows by symmetry.

As a side-note, we mention that if eigenvalue $1$ has multiplicity more than one, then all the above discussed minor matrices have determinant $\equiv 0$. Following the above argument shows that in such cases the coin must be a direct sum of one-dimensional trapping coins.

Finally, note that our proof of the first case directly generalizes to higher dimensions, and we think that it should also be possible to handle the degenerate second case in greater generality. This suggests that for higher dimensional square lattices the stationary eigenstates can also be confined into $2\times 2 \times \ldots \times 2$ regions for the basic quantum walk where the displacements in all directions are by $\pm 1$. A very recent extension~\cite{Konno2020} of earlier work~\cite{komatsu2017} about the stationary states of the Grover walk on $\mathds{Z}^d$ also supports this conjecture.

%\section*{References}
\bibliographystyle{apsrev4-2}
\bibliography{biblio}

\end{document}